\documentclass[prd,aps,showpacs,floats,letterpaper,floatfix,groupedaddress,eqsecnum,amssymb,amsmath]{revtex4}

\usepackage{graphicx,epsfig}
\usepackage{bm}
\usepackage{amsmath}
\usepackage{dcolumn,epsfig}

\def\be{\begin{equation}}
\def\ee{\end{equation}}
\def\beq{\begin{eqnarray}}
\def\eeq{\end{eqnarray}}
\def\ii{{\rm i}}

\begin{document}

\centerline{}
\title{Eigenvalues and eigenfunctions of spin-weighted spheroidal
harmonics \\ in four and higher dimensions}
\author{Emanuele Berti}
\email{berti@wugrav.wustl.edu} \affiliation{McDonnell Center for the
Space Sciences, Department of Physics, Washington University, St.
Louis, Missouri 63130, USA}
\author{Vitor Cardoso}
\email{vcardoso@phy.olemiss.edu} \affiliation{Dept. of Physics and Astronomy, The University of
Mississippi, University, MS 38677-1848, USA \footnote{Also at Centro de F\'{\i}sica Computacional,
Universidade de Coimbra, P-3004-516 Coimbra, Portugal}}
\author{Marc Casals}
\email{marc.casals@ucd.ie}
\affiliation{School of Mathematical Sciences, University College
Dublin, Belfield, Dublin 4, Ireland}

\date{\today}

\begin{abstract}
Spin-weighted spheroidal harmonics are useful in a variety of physical
situations, including light scattering, nuclear modeling, signal
processing, electromagnetic wave propagation, black hole perturbation
theory in four and higher dimensions, quantum field theory in curved
space-time and studies of D-branes. We first review analytic and
numerical calculations of their eigenvalues and eigenfunctions in four
dimensions, filling gaps in the existing literature when
necessary. Then we compute the angular dependence of the spin-weighted
spheroidal harmonics corresponding to slowly-damped quasinormal mode
frequencies of the Kerr black hole, providing numerical tables and
approximate formulas for their scalar products. Finally we present an
exhaustive analytic and numerical study of scalar spheroidal harmonics
in ($n+4$) dimensions.
\end{abstract}


\maketitle


\section{Introduction}

Spheroidal harmonics are special functions that play an important role
in mathematical physics. The simplest prototype are the {\it scalar
spheroidal harmonics} ${}_0S_{lm}$, obeying the differential equation
\begin{equation}
\left [(1-x^2){}_0S_{lm,x} \right]_{,x} +\left[ (c x)^2+{}_0A_{lm}
-{m^2\over 1-x^2} \right] {}_0S_{lm}=0\,.\label{scalarswsh}
\end{equation}
where $l(=0,~1,~2\dots)$ and $m$ are integers, $|m|\leq l$ and
${}_0A_{lm}$ is an angular eigenvalue to be found by imposing
appropriate boundary conditions. The parameter $c$ is, in general,
complex. When $c\in \mathbb{R}$ the eigenfunctions are called {\it
oblate}; when $c$ takes on pure imaginary values ($c=\ii c_I$ with
$c_I\in \mathbb{R}$) the eigenfunctions are called {\it prolate}
\cite{flammer}.  Scalar spheroidal harmonics show up in a variety of
physical situations ranging from light scattering \cite{lightscatt} to
nuclear modeling \cite{nuclear}, signal processing and electromagnetic
wave propagation \cite{electromagnetic}. There is a solid body of work
on scalar spheroidal harmonics, some classic references being Stratton
{\it et al.} \cite{stratton}, Meixner and Sch\"{a}fke \cite{meixner}
and Flammer \cite{flammer} (see also \cite{oguchi,li,barrowes,falloon}
for more recent developments on the subject).

Generalized (four-dimensional) {\it spin-weighted spheroidal
harmonics} (SWSHs) were first defined by Teukolsky \cite{teukolsky} in
the context of black hole physics.  They result from the separation of
angular variables in the equations describing the propagation of a
spin-$s$ field in a rotating (Kerr) black hole background. Using the
Kinnersley tetrad and Boyer-Lindquist coordinates, SWSHs satisfy a
generalized version of Eq.~(\ref{scalarswsh}) [see
Eq.~(\ref{angularwaveeq}) below]. Now the angular separation constant,
denoted by $_{s}A_{l m}$, depends also on a spin-weight parameter
$s=0,\pm 1/2,\pm 1,\pm 2$ when we consider scalar, neutrino,
electromagnetic or gravitational perturbations, respectively. When
$s=0$ SWSHs reduce to ordinary, scalar spheroidal harmonics.  When
$c=0$ they reduce to spin-weighted {\it spherical} harmonics
\cite{goldberg}, which have innumerable applications in physics (see
eg.~\cite{wjv} for an application to the analysis of anisotropies in
the cosmic microwave background). The ordinary spherical harmonics are
spin-weighted spherical harmonics with $s=0$.

Describing the angular dependence of a spin-$s$ perturbing field in a
Kerr black hole background, SWSHs and their eigenvalues find
application in many theoretical studies of black hole physics.

An important application concerns quasinormal modes of Kerr black
holes \cite{kokkotas}. The damped oscillation frequencies of Kerr
black holes, $\omega$, are uniquely determined by the black hole's
mass $M$ and specific angular momentum $a$. The calculation of
quasinormal frequencies reduces to the solution of a coupled system of
differential equations. One equation belongs to the class of
generalized spheroidal wave equations, and describes the radial
dependence of the perturbations; the other is a SWSH with $c=a\omega$,
describing the angular dependence
\cite{leaver,leaver2,hisashi,bertikerr2,BCY,milan}.  For this reason,
a detailed knowledge of the numerical and analytic properties of SWSHs
is necessary to compute black hole quasinormal modes.  Recent
conjectures suggest a relation between highly damped quasinormal modes
and black hole area quantization \cite{hod,dreyer}. Verifying these
conjectures for Kerr black holes calls for a calculation of the
eigenvalues ${}_sA_{lm}$ corresponding to values of $c$ with imaginary
part much larger than the real part, $|c_I|\gg |c_R|$.  More
phenomenologically, an important astrophysical problem is the
determination of black hole parameters from gravitational wave
observations \cite{echeverria,bertiringdown}. Gravitational waves
emitted by the oscillations of a rotating black hole can be described
by a superposition of quasinormal modes. The Laser Interferometer
Space Antenna (LISA) has the potential to detect these waves with
large signal-to-noise ratio \cite{FH}. Given a detection, we would
like to extract as much information as possible about the
source. Ideally, we would like to determine $M$, $a$, the source
location and the black hole's spin orientation from the observed
waveform. An investigation of this issue requires the calculation of
``scalar products'' between different quasinormal modes, and in
particular between the SWSHs describing their angular dependence
\cite{bertiringdown}.

SWSHs also find application in quantum field theory in curved
space-time. This theory is plagued with ultraviolet divergences: in
particular, the expectation value of the stress-energy tensor, which
is the crucial quantity that generates space-time curvature via the
semiclassical Einstein's equation, suffers from ultraviolet
divergences and must be renormalized.  This is usually a very
difficult endeavour, particularly in the case of non-spherically
symmetric space-times, which have recently been studied within this
framework in \cite{Hawking4Dim}.  An understanding of the
high-frequency behaviour (both of the radial and angular parts) of the
matter fields is essential in order to renormalize the various
quantities in the theory \cite{CO}.

Many theoretical scenarios rely on our Universe being
($4+n$)-dimensional, with the $n$ extra dimensions compactified on
some small scale \cite{hamed}. The study of higher-dimensional black
hole geometries, such as the Myers-Perry solution \cite{myersperry},
requires the introduction of {\it higher-dimensional spheroidal
harmonics} (HSHs), that we shall define in Sec.~\ref{sec:ddim}
\cite{ida,frolov,vasudevan,css}. TeV-scale gravity scenarios allow for
the possibility that rotating, higher-dimensional mini-black holes are
produced in particle accelerators such as CERN's Large Hadron
Collider. These higher-dimensional, rotating black holes should
evaporate by emission of Hawking radiation
\cite{frolov,HawkingHigherDim}, and both SWSHs and HSHs are useful for
a quantitative study of the evaporation process on the ($3+1$)-brane
and in the ($4+n$)-bulk, respectively.  HSHs also show up in recent
studies of smooth geometries in the D1-D5 system
\cite{mathur}. Separating the Klein-Gordon equation leads again to
five-dimensional spheroidal harmonics (even though they were not
identified as such in \cite{mathur}).

There are no in-depth investigations of HSHs to date. On the other
hand, the properties of SWSHs were (at least partially) investigated
as soon as they were introduced. Press and Teukolsky \cite{PT}
provided a polynomial fit in $c$ of the eigenvalues ${}_sA_{lm}$,
which is valid up to $c \sim 3$.  A formal perturbation expansion in
powers of $c$ was carried out by Fackerell and Crossman
\cite{fackerell} (see also \cite{seidel}, where some typos were
corrected). An expansion for large real values of $c$ has long been
known for scalar spheroidal harmonics
\cite{meixner,flammer,oguchi}. For general spin-$s$ fields, the first
attempt to find a large-$c$ expansion was made by Breuer, Ryan and
Waller \cite{breuer,breuerbook}. Their analysis, which was partially
flawed and incomplete, has recently been revisited and corrected by
Casals and Ottewill \cite{CO}. An expansion of the eigenvalue for
large, pure-imaginary $c$ is known for scalar spheroidal harmonics
\cite{meixner,flammer,oguchi}, and was partially studied in
\cite{breuerbook,BCY} for general spin-$s$ fields. However, as we
already mentioned some of the studies for general spin-$s$ were
flawed, and we still lack a unified picture of the situation.

The main purpose of this paper is to provide a complete analysis of
general (four-dimensional) SWSHs, and to extend this understanding to
HSHs. The plan of the paper is as follows.

Sec.~\ref{sec:4dim} contains analytic and numerical results for
four-dimensional SWSHs. In Sec.~\ref{eqns} we define SWSHs and present
a series solution first obtained by Leaver \cite{leaver}. In
Sec.~\ref{numerical} we explain how this series solution provides us
with a simple numerical algorithm to compute eigenvalues and
eigenfunctions, and discuss alternative methods we used to check our
results. In Sec.~\ref{4Dsmallaw} we give the series expansion of the
eigenvalues for small $c$. In Sec.~\ref{subsec:4SWSHlarge,real} and
\ref{prolate-large} we present analytic results for large values of
the argument in the oblate and prolate cases, respectively,
supplementing these analytic expansions by numerical
calculations. Sec.~\ref{EFnum} contains a numerical calculation of the
prolate eigenfunctions. In Sec.~\ref{KerrQNMs} we compute numerically
the SWSHs at the slowly damped quasinormal frequencies of a Kerr black
hole, providing numerical tables and a simple analytic approximation
of their scalar products. While Secs.~\ref{eqns}, \ref{numerical},
\ref{4Dsmallaw} and \ref{subsec:4SWSHlarge,real} are mostly review
material, the results in Secs.~\ref{prolate-large}, \ref{EFnum} and
\ref{KerrQNMs} are new.

Sec.~\ref{sec:ddim} is devoted to analytic and numerical calculations
of the eigenvalues and eigenfunctions for HSHs. In Sec.~\ref{idaop} we
review a series representation due to Ida {\it et al.} \cite{ida},
illustrating the corresponding computational algorithm for the
eigenvalues and eigenfunctions. The results in the following Sections
are entirely new. In Sec.~\ref{sec:ddimexp-small} we give a series
expansion of the eigenvalues for small $c$.
Secs.~\ref{sec:ddimexp-large} and \ref{sec:ddimexp-large,prolate}
present series expansions for large values of the argument in the
oblate and prolate cases, respectively, and back them up by numerical
calculations.  We conclude with a summary of relevant results and open
problems.

Other original results are contained in the Appendices.  In Appendix
\ref{sec:A1} we give an exact analytic solution of the HSH equation
for $c=0$, and use it to determine the number of zeros of the
eigenfunctions in the region of physical interest.  Appendix
\ref{sec:A2} provides a simple analytic solution of the HSH equation
for special values of the parameters.

\section{Four-dimensional spin-weighted spheroidal harmonics}
\label{sec:4dim}

\subsection{Series solution}
\label{eqns}

Using the Kinnersley tetrad and Boyer-Lindquist coordinates, the
angular equation definining SWSHs results from the separation of the
equations describing propagation of a spin-$s$ field in the Kerr
background \cite{teukolsky}:
\begin{equation}
\label{angularwaveeq}
\left [(1-x^2){}_sS_{lm,x} \right]_{,x} +\left[
(cx)^2-2c sx+s+{}_{s}A_{lm}
 -{(m+sx)^2\over 1-x^2}
\right] {}_sS_{lm}=0\,,
\end{equation}
where $x\equiv \cos\theta$ and $\theta$ is the Boyer-Lindquist polar
angle.

The angular separation constant ${}_{s}A_{lm}$ and the SWSHs
${}_sS_{lm}$ are, in general, complex. They take on real values only
in the oblate case ($c\in \mathbb{R}$) or, alternatively, in the
prolate case ($c=\ii c_I$ pure-imaginary) with $s=0$. In the limit
$c\to 0$ the angular separation constant can be determined
analytically:
\begin{equation}\label{schwlim}
{}_sA_{lm}=l(l+1)-s(s+1)\,.
\end{equation}
Some useful symmetry properties hold (see eg. \cite{leaver}):

(i) Given eigenvalues for (say) positive $m$, those for negative $m$
are readily obtained by complex conjugation:
\begin{equation}\label{conj}
{}_sA_{lm}={}_sA_{l-m}^*\,;
\end{equation}

(ii) Given eigenvalues for (say) negative $s$, those for positive $s$
are given by
\begin{equation}\label{negs}
_{-s}A_{lm}={}_sA_{lm}+2s\,.
\end{equation}
Exploiting these symmetries, in our numerical calculations we only
consider $s\leq 0$ and $m\geq 0$.

(iii) Let us define $\rho\equiv \ii c$. If $\rho$ and $_{-s}A_{lm}$
corresponds to a solution for given $(s,~l,~m)$, then another solution
can be obtained by the following replacements: $m\to -m$, $\rho\to
\rho^*$, $_{-s}A_{lm}\to _{-s}A_{l-m}^*$.

(iv) In the prolate case, if ${}_sS_{lm}$ is a solution with
eigenvalue ${}_sA_{lm}$ and $c_I>0$, then ${}_sS_{lm}^*$ is a solution
with eigenvalue ${}_sA^*_{lm}$ and $c_I<0$.

Leaver found the following series solution for the angular
eigenfunctions \cite{leaver}:
\begin{equation}\label{leavers}
{}_sS_{lm}(x)=e^{c x}\left(1+x\right)^{k_-}\left(1-x\right)^{k_+}
\sum_{p=0}^\infty a_p(1+x)^p\,,
\end{equation}
where $k_{\pm}\equiv |m\pm s|/2$. The expansion coefficients $a_p$ are
obtained from the three term recursion relation
\begin{eqnarray}\label{recur}
\alpha_0 a_{1}+\beta_0 a_{0}&=&0\,,\\
\alpha_p a_{p+1}+\beta_p a_{p}+\gamma_p a_{p-1}&=&0\,, \qquad p=1,2\dots \label{recur4D}
\end{eqnarray}
with
\begin{eqnarray} \label{recur2}
\alpha_p&=&-2(p+1)(p+2k_-+1)\,,\\
\beta_p&=&p(p-1)+2p(k_-+k_++1-2c)\nonumber\\
&-&\left[
2c\left(2k_-+s+1\right)-\left(k_-+k_+\right)\left(k_-+k_++1\right)
\right]-\left[c^2+s(s+1)+{}_sA_{lm}\right]\,,\nonumber \label{bet4D}\\
\gamma_p&=&2c\left(p+k_-+k_++s\right)\,.\nonumber
\end{eqnarray}

Given a (generally complex) argument $c$, the separation constant
${}_sA_{lm}$ can be obtained solving numerically the continued
fraction equation
\begin{eqnarray}\label{cfeq}
\beta_0-
{\alpha_0\gamma_1\over\beta_1-}
{\alpha_1\gamma_2\over\beta_2-}
{\alpha_2\gamma_3\over\beta_3-}
...=0 \,,
\end{eqnarray}
or any of its inversions \cite{leaver}.

\subsection{Numerical calculation of the eigenvalues and eigenfunctions}
\label{numerical}

Flammer's classical reference \cite{flammer} is largely dedicated to
tabulating eigenvalues and eigenfunctions of scalar spheroidal
harmonics. The main limitation of Flammer's impressive work is that
his tables only deal with the prolate (pure imaginary frequency) and
oblate (pure real frequency) cases. The general case of complex
frequencies (which is of interest, for example, in the calculation of
the angular dependence of quasinormal modes) is not covered. A
pioneering numerical work in this sense was carried out by Oguchi
\cite{oguchi}, who computed angular eigenvalues for complex values of
$c$ and $s=0$. Quite recently this topic received more attention. Li
{\it et al.}  published a useful review of numerical methods to
compute eigenvalues and eigenfunctions for $s=0$ \cite{li}.  Falloon
{\it et al.}  developed a {\it Mathematica} notebook to compute $s=0$
harmonics for general complex values of the frequency \cite{falloon}.
Finally, Barrowes {\it et al.} provided a rapid and accurate method to
calculate the prolate and oblate scalar spheroidal wave functions and
their eigenvalues for complex frequencies in the limit $|c|\to \infty$
\cite{barrowes}.

In comparison, numerical calculations of the eigenfunctions and
eigenvalues for general spin $s$ are scarce.  To make things worse,
until recently the few analytic predictions were contradictory (the
situation for large {\it real} frequencies has finally been clarified
in \cite{CO}, while large {\it pure imaginary} frequencies were
considered in \cite{BCY}).

Leaver's solution gives a simple and practical algorithm for the
numerical calculation of eigenvalues and eigenfunctions. Start from
the known analytic eigenvalue for $c=0$, Eq.~(\ref{schwlim}). Use this
as an initial guess, increase the value of $c$ and solve numerically
Eq.~(\ref{cfeq}) to get the eigenvalue for $c\neq 0$. Once the
eigenvalue is known, compute any number of series coefficients $a_p$
using the recursion relation, and plug them into the series solution
(\ref{leavers}) to get the corresponding eigenfunction to any required
precision. In our numerical calculations we truncate the series at
$p=p_{\rm max}=1000$ (but usually an excellent approximation to the
eigenfunction can be obtained keeping only $\sim 10$ terms).

This algorithm only determines the eigenfunction up to a normalization
constant, which can easily be fixed by imposing the normalization
condition
\be
\int_{-1}^{1} |{}_sS_{lm}(x)|^2 dx=1\,.
\ee

We independently checked our numerical results for the eigenvalues and
eigenfunctions using different methods. These methods are variants of
those described in detail in \cite{CO}, where they were used to deal
with oblate SWSHs, so we only describe them briefly.  To check the
eigenvalues from Leaver's continued fractions we adapt the procedure
derived by Sasaki and Nakamura \cite{Sasa&Naka} for $s=-2$ and $c\in
\mathbb{R}$.  The idea is to approximate the angular differential
equation by a difference equation, which can be written in matrix
form.  The eigenvalue is then obtained by imposing the determinant of
this (tridiagonal) matrix to be zero.  The Sasaki-Nakamura method
faces a numerical difficulty as the discretization grid is refined:
the value of the determinant becomes very large, making it
increasingly difficult to obtain the eigenfunctions. Therefore, in
order to find the eigenfunctions we complement the Sasaki-Nakamura
analysis by an adaptation of the shooting method introduced in
\cite{NumRec} for the special case of (scalar) spheroidal harmonics
with real frequency.  We start the integration of the angular equation
with arbitrary values for (i) the eigenvalue, (ii) the angular
function at $x=-1$.  The eigenvalue results from imposing the
derivative of ${}_sS_{lm}$ near $x=+1$, as found by numerical
integration, to agree with a certain analytic approximation of the
derivative, coming from a power series expansion around $x=+1$.
In \cite{CO}, for large real frequency it was sometimes necessary to
look for an extreme (rather than a zero) of either the determinant or
the difference of derivatives near $x=+1$.  That is because in the
oblate case the eigenvalues for some modes ``pair up'' for large
frequency.
This degeneracy does not occur for large and {\it pure-imaginary}
frequency (prolate case).  On the downside, when using the
``shooting'' method we must now find the zero of a complex
function. We locate this zero using Newton's method \cite{NumRec},
which is globally convergent.

\begin{figure}[hbt]
\centerline{
\includegraphics[width=9cm,angle=0]{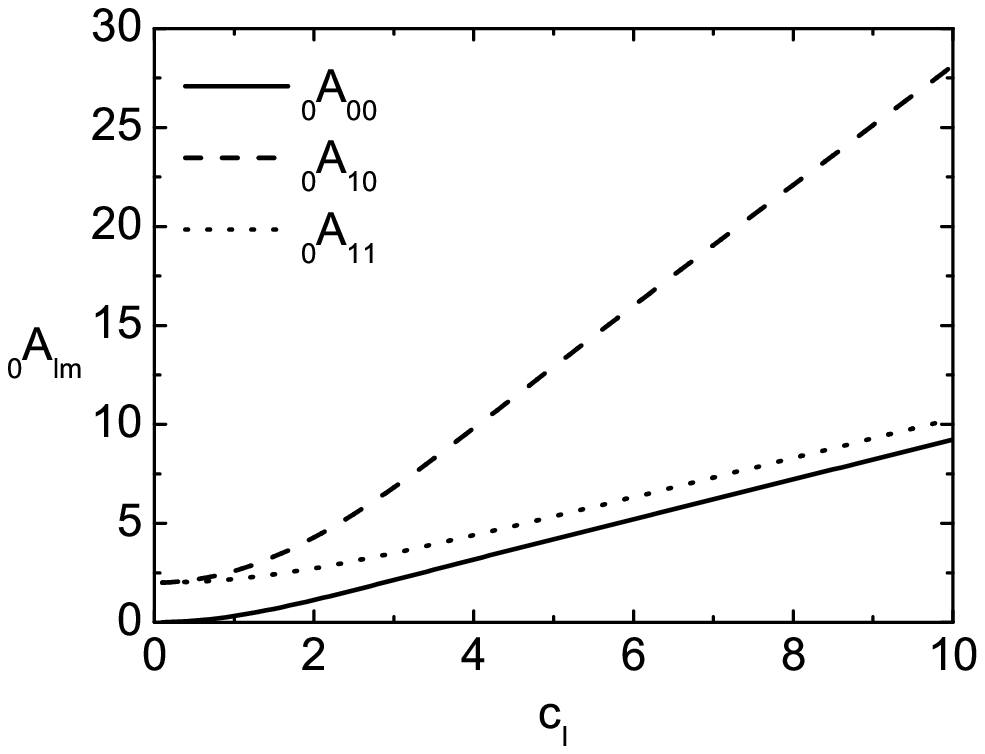}
\includegraphics[width=9cm,angle=0]{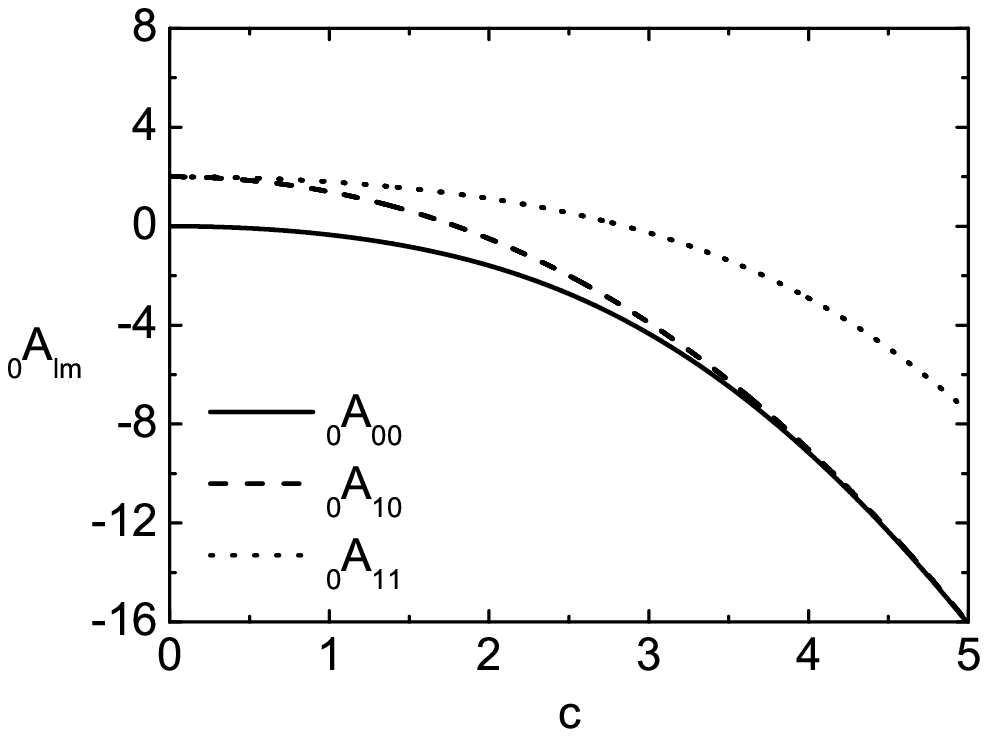}
}
\caption{Angular eigenvalue $_{0}A_{lm}$ as a function of $c_I$
(prolate case) or $c$ (oblate case), for selected values of $l$ and
$m$.}
\label{fig:1}
\end{figure}

As a first example of the application of Leaver's method, in
Fig.~\ref{fig:1} we show some prolate and oblate eigenvalues for
$s=0$. Progressively increasing the inversion index of the continued
fraction \cite{leaver} we can track the eigenvalues (which in this
special case are real) up to quite large values of $c$.  Our numerical
code passed a number of tests. We first verified that our results are
in agreement with Tables 10-12 of \cite{flammer} in the prolate case
and Tables 130-132 of \cite{flammer} in the oblate case. For complex
$c$ we were able to reproduce Table 2 in \cite{oguchi}. As a final
sanity check, we verified in a few representative cases that the {\it
Mathematica} notebooks presented in \cite{li, falloon} are also in
agreement with our code.

Notice that prolate and oblate eigenvalues coincide when $c\to 0$, but
their behavior for large frequencies is markedly different. Prolate
eigenvalues grow {\it linearly} with $|c_I|$ for large $|c_I|$, while
oblate eigenvalues are proportional to $-c^2$ for large $c$. In the
following Section we derive analytically asymptotic expansions in the
limit of small $c$, large real $c$ and large pure-imaginary $c$.

\subsection{Small-$c$ expansion}
\label{4Dsmallaw}

The analytic properties of the eigenvalues and eigenfunctions of the
SWSH equation (\ref{angularwaveeq}) have been studied by many authors
\cite{meixner,flammer,BCY,breuerbook,PT,fackerell,breuer,seidel,CO}.
Here and in the following Sections we present for the first time a
compact and complete summary of the analytic results for the
eigenvalues in various asymptotic regimes.

The small-$c$ expansion was worked out many years ago for the scalar
case $s=0$ (see eg. \cite{flammer}). The general $s$ case was
considered by Press and Teukolsky \cite{PT}, Fackerell and Crossman
\cite{fackerell} and Seidel \cite{seidel}. Here we present an
alternative and simpler derivation of their results, which can easily
be extended to the higher-dimensional case (see
Sec.~\ref{sec:ddimexp-small}). Our method is directly based on the
continued fraction representation. For $c=0$, the eigenvalue
${}_sA_{lm}$ is explicitly determined from the requirement that the
series expansion has a finite number of terms, since otherwise it is
divergent \cite{leaver}. Taking $c=0$ in (\ref{bet4D}) and imposing
$\beta_r=0$ for some integer $r\geq 0$ we obtain
\be \label{A,c=}
{}_sA_{lm}=(r+|m|)(r+|m|+1)-s(s+1)\,.
\ee
Appendix \ref{sec:A1} gives an alternative derivation of this result
in a more general framework. Setting $r=l-|m|$ we recover the result
(\ref{schwlim}); furthermore, the condition $r\geq 0$ implies the
well-known constraint on the angular quantum numbers, $l\geq |m|$.

For finite but small $c$ one can use standard perturbation theory to
express the separation constant as a power series in $c$. The standard
procedure is somewhat involved \cite{fackerell,seidel}, but it can be
greatly simplified as follows. For $c=0$ we know the recursion
relation (\ref{recur}) has a finite number of terms $r$. Therefore,
when expanding around $c=0$ we find it convenient (although strictly
not necessary) to use the $r$-th inversion of Eq.~(\ref{cfeq}):
\be \beta _r-\frac{\alpha_{r-1}\gamma _{r}}{\beta
_{r-1}}
\frac{\alpha_{r-2}\gamma_{r-1}}{\beta_{r-2}-}\dots
\frac{\alpha _0 \gamma _1}{\beta_0}=\frac{\alpha _r
\gamma _{r+1}}{\beta _{r+1}-}
\frac{\alpha _{r+1}\gamma _{r+2}}{\beta_{r+2}-}\dots\, \label{cont2}\ee
Now we expand the separation constant as a Taylor series:
\be
{}_sA_{lm}=\sum_{p=0}^{\infty} f_pc^p\,,
\label{expansion}
\ee
where $f_0=l(l+1)-s(s+1)$. Plugging (\ref{expansion}) into
Eq.~(\ref{cont2}), expanding the resulting expression in powers of $c$
and defining
\be \frac{1}{2}(\alpha+\beta)={\rm max}(|m|,|s|)\,\,,\,\,\,\,\frac{1}{2}(\alpha-\beta)=\frac{ms}{{\rm
max}(|m|,|s|)}\,, \ee
\be
h(l)=\frac{\left[l^2-\frac{1}{4}(\alpha+\beta)^2\right]
\left[l^2-\frac{1}{4}(\alpha-\beta)^2\right] (l^2-s^2)}
{2\left(l-\frac{1}{2}\right)l^3\left(l+\frac{1}{2}\right)}\,,
\ee
we have for the first coefficients
\begin{subequations}
\beq f_0&=&l(l+1)-s(s+1) \label{4dsmall1}\,,\\
f_1&=&-\frac{2ms^2}{l(l+1)}\,,\\
f_2&=&h(l+1)-h(l)-1\,,\\
f_3&=&\frac{2h(l)ms^2}{(l-1)l^2(l+1)}-\frac{2h(l+1)ms^2}{l(l+1)^2(l+2)}\,,\\
f_4&=&m^2s^4\left
(\frac{4h(l+1)}{l^2(l+1)^4(l+2)^2}-\frac{4h(l)}{(l-1)^2l^4(l+1)^2}\right
)-\frac{(l+2)h(l+1)h(l+2)}{2(l+1)(2l+3)}+\nonumber
\\ &&
\frac{h^2(l+1)}{2(l+1)}+\frac{h(l)h(l+1)}{2l^2+2l}-\frac{h^2(l)}{2l}+\frac{(l-1)h(l-1)h(l)}{4l^2-2l}\,,\\
f_5&=&m^3s^6\left (\frac{8h(l)}{l^6(l+1)^3(l-1)^3}-\frac{8h(l+1)}{l^3(l+1)^6(l+2)^3}\right )+\nonumber
\\
& & ms^2h(l)\left ( -\frac{
h(l+1)(7l^2+7l+4)}{l^3(l+2)(l+1)^3(l-1)}-\frac{
h(l-1)(3l-4)}{l^3(l+1)(2l-1)(l-2)} \right )+\nonumber \\
& & ms^2 \left (\frac{(3l+7) h(l+1)h(l+2)}{l(l+1)^3(l+3)(2l+3)}-\frac{3h^2(l+1)}{l(l+1)^3(l+2)}
+\frac{3 h^2(l)}{l^3(l-1)(l+1)} \right )\,, \\
f_6&=&\frac{16m^4s^8}{l^4(l+1)^4}\left
(\frac{h(l+1)}{(l+1)^4(l+2)^4}-\frac{h(l)}{l^4(l-1)^4}\right )+
\nonumber \\
& &\frac{4m^2s^4}{l^2(l+1)^2}\left
(-\frac{(3l^2+14l+17)h(l+1)h(l+2)}{(l+1)^3(l+2)(l+3)^2(2l+3)}+\frac{3h^2(l+1)}{(l+1)^3(l+2)^2}
-\frac{3h^2(l)}{l^3(l-1)^2}\right )+
\nonumber \\
& & \frac{4m^2s^4}{l^2(l+1)^2}\left ( \frac{(11l^4+22l^3+31l^2+20l+6)h(l)h(l+1)}{l^3(l-1)^2(l+1)^3(l+2)^2}+
\frac{(3l^2-8l+6)h(l-1)h(l)}{l^3(l-2)^2(l-1)(2l-1)} \right
)+\nonumber \\
& & \frac{h(l+1)h(l+2)}{4(l+1)(2l+3)^2}\left (
\frac{(l+3)h(l+3)}{3}+\frac{l+2}{l+1}\left (
(l+2)h(l+2)-(7l+10)h(l+1)+\frac{(3l^2+2l-3)h(l)}{l}
\right ) \right )+\nonumber \\
& & \frac{h^3(l+1)}{2(l+1)^2}-\frac{h^3(l)}{2l^2}
+\frac{h(l)h(l+1)}{4l^2(l+1)^2}\left (
(2l^2+4l+3)h(l)-(2l^2+1)h(l+1)-\frac{(l^2-1)(3l^2+4l-2)h(l-1)}{(2l-1)^2}
\right )+\nonumber \\
& & \frac{h(l-1)h(l)}{4l^2(2l-1)^2}\left ((l-1)(7l-3)h(l)-(l-1)^2h(l-1)-\frac{l(l-2)h(l-2)}{3} \right
)\,, \label{4dsmall2} 
\eeq
\end{subequations}
in agreement with \cite{seidel}. Notice that our separation constant
${}_sA_{lm}$ differs from the separation constant ${}_sE_{lm}$ of
Refs.~\cite{fackerell,seidel}: ${}_sA_{lm}={}_sE_{lm}-s(s+1)$.

\subsection{Large and real $c$ (oblate case)}
\label{subsec:4SWSHlarge,real}

For $s=0$, analytic power-series expansions for large (pure real and
pure imaginary) values of $c$ have long been known
\cite{meixner,flammer}. In this Subsection and in the next we review
and extend known analytic results for general values of the
spin-weight $s$.  Large-$c$ expansions of the eigenvalues are usually
based on the following idea.  First, an asymptotic approximation of
the solution is found.  An asymptotic expansion of the eigenvalue is
then obtained by matching the number of zeros of the asymptotic
solution with the (known) number of zeros of the exact solution (as
determined, eg., in \cite{breuer}). This matching relies on the
solution being analytic in $c$.

Breuer, Ryan and Waller \cite{breuer} perform an asymptotic analysis
for large, real $c$ (and fixed $m$) of the SWSH equation. Their
analysis is based on similar analyses in the $s=0$ limit
\cite{meixner,flammer} and on a partially-flawed previous study for
general spin \cite{breuerbook}. It yields the following asymptotic
expansion for the separation constant:

\begin{equation} \label{eq:series E for large w}
\begin{aligned}
{}_{s}A_{lm}&=-c^{2}+2{}_{s}q_{lm}c-\frac{1}{2}\left[{}_{s}q_{lm}^2-m^2+2s+1\right]
+\frac{1}{c}A_1-
\\ & -
\frac{1}{64c^2}\left[5{}_{s}q_{lm}^4-\left(6m^2-10\right){}_{s}q_{lm}^2+m^4-2m^2-4s^2\left({}_{s}q_{lm}^2-m^2-1\right)+1\right]+
\\ & +
\frac{1}{c^3}A_3+{\cal O}\left(1/c^{4}\right)
\end{aligned}
\end{equation}
where
\begin{equation}
\begin{aligned}
A_1&=-\frac{1}{8}\left[{}_{s}q_{lm}^3-m^2{}_{s}q_{lm}+{}_{s}q_{lm}-s^2\left({}_{s}q_{lm}+m\right)\right] \\
A_3&=\frac{1}{512} \Bigg\{\frac{1}{64}\Big[
\left({}_{s}q_{lm}-m-1+2s\right)\left({}_{s}q_{lm}-m-1-2s\right)
\left({}_{s}q_{lm}-m-2s-3\right) \times \\ & \times
\left({}_{s}q_{lm}-m+2s-3\right)
\left({}_{s}q_{lm}+m-3\right)^2\left({}_{s}q_{lm}+m-1\right)^2-
\\ & -
\left({}_{s}q_{lm}-m-2s+1\right) \left({}_{s}q_{lm}-m+2s+1\right)
\left({}_{s}q_{lm}-m-2s+3\right) \left({}_{s}q_{lm}-m+2s+3\right)
\times \\ & \times \left({}_{s}q_{lm}+m+1\right)^2
\left({}_{s}q_{lm}+m+3\right)^2\Big]+
\\ & +
2\Big[\left({}_{s}q_{lm}-m+2s-1\right)\left({}_{s}q_{lm}-m-2s-1\right)\left({}_{s}q_{lm}-1\right)^2
\left({}_{s}q_{lm}+m-1\right)^2-
\\ & -
\left({}_{s}q_{lm}+m+1\right)^2\left({}_{s}q_{lm}+1\right)^2\left({}_{s}q_{lm}-m-2s+1\right)
\left({}_{s}q_{lm}-m+2s+1\right)\Big]-
\\ & -
2A_1\Big[\left({}_{s}q_{lm}-m+2s-1\right)
\left({}_{s}q_{lm}-m-2s-1\right)\left({}_{s}q_{lm}+m-1\right)^2+
\\ & +
\left({}_{s}q_{lm}+m+1\right)^2\left({}_{s}q_{lm}-m-2s+1\right)\left({}_{s}q_{lm}-m+2s+1\right)\Big]
\Bigg\}
\end{aligned}
\end{equation}

Unfortunately, Ref.~\cite{breuer} left the parameter ${}_{s}q_{lm}$
undetermined. The recent work by Casals and Ottewill \cite{CO}
determines ${}_{s}q_{lm}$ through a complete asymptotic analysis for
large, real $c$ (and fixed $m$). First they find an asymptotic
expansion for the angular solution valid everywhere in $x\in
[-1,+1]$. Then they determine ${}_{s}q_{lm}$ by imposing the
asymptotic angular solution to have the right number of zeros. This
could not be done successfully by Breuer, Ryan and Waller as their
asymptotic angular solution was not valid near the origin $x=0$, where
the solution may have one zero. The value of the parameter
${}_{s}q_{lm}$ obtained in \cite{CO} is:
\begin{subequations}
\label{eq:val. of q}
\begin{align}
{}_sq_{lm}&= l-{\rm max}(|m|,|s|) +\frac{(|m+s |+|m-s |)}{2}+1-z_{0}
\quad
&& \text{if} \quad l \geq {\rm max}({}_{s}l_{m},{}_{-s}l_{m})
\label{eq:1st q} \\
{}_sq_{lm}&= 2\left(l-{\rm max}(|m|,|s|)\right)+|m\mp s |\mp s+1
\quad && \text{if} \quad l<{}_{\pm s}l_{m}  \label{eq:3rd q}
\end{align}
\end{subequations}
where ${}_sl_{m}\equiv {\rm max}(|m|,|s|)+(|m+s |-|m-s |)/2+s$, and
the variable
\begin{equation} \label{eq:z0=0,1 if S has zero at x=0 or not}
z_{0}=
\begin{cases}
0 & \text{if} \quad l-{}_sl_{m} \quad \text{even}   \\
1 & \text{if} \quad l-{}_sl_{m} \quad \text{odd}
\end{cases}
\end{equation}
gives the number of zeros of the asymptotic angular solution at $x=0$.

\begin{figure}[hbt]
\centerline{
\includegraphics[width=9cm,angle=0]{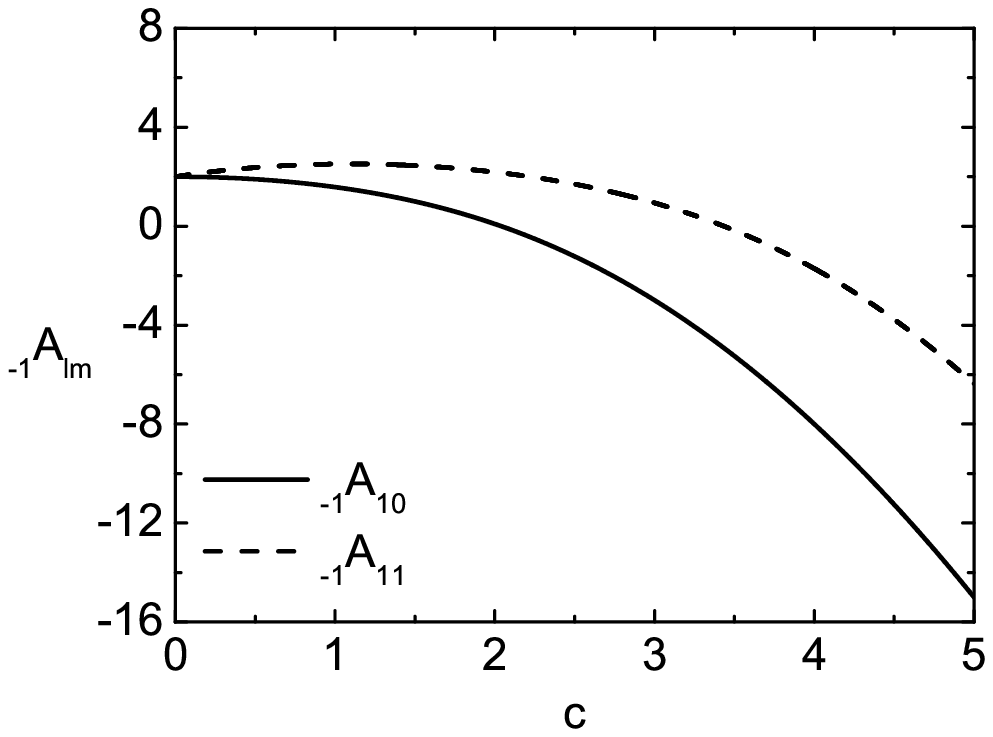}
\includegraphics[width=9cm,angle=0]{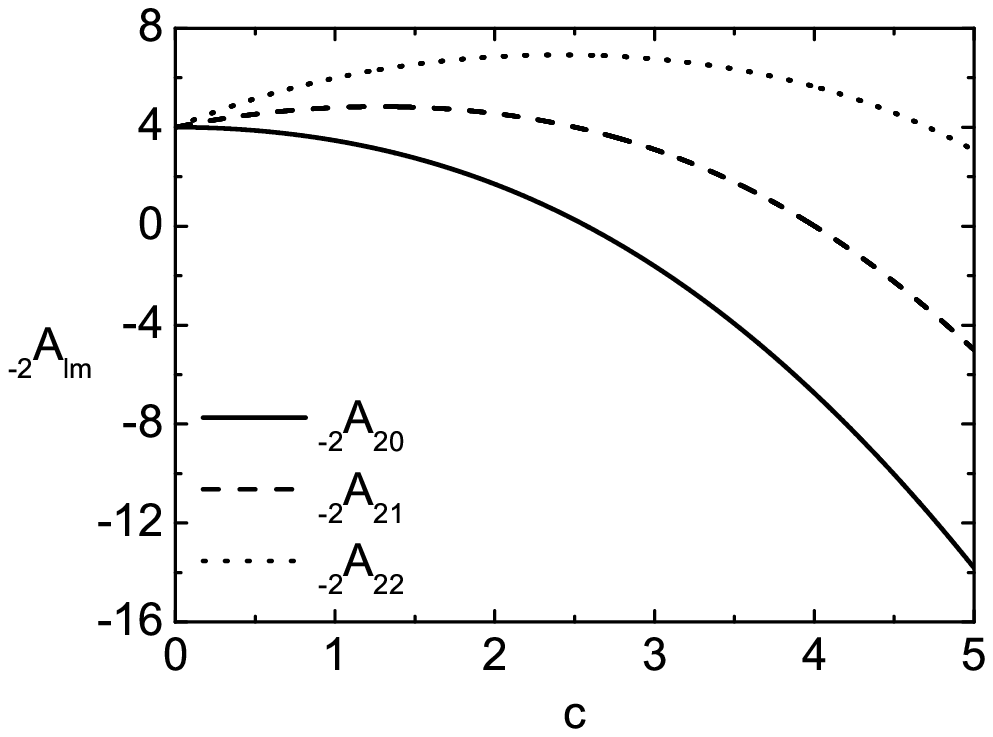}}
\caption{Angular eigenvalue for the lowest radiatable multipoles with
$s=-1$ (left) and $s=-2$ (right) in the oblate case.}
\label{fig:2}
\end{figure}

In Fig.~\ref{fig:2} we use Leaver's method to give numerical support
to the analytic results, computing selected oblate eigenvalues for
$s\neq 0$. Fig.~\ref{fig:2} only shows the lowest radiatable
multipoles (that is, the lowest values of $l$ such that $l\geq |s|$),
which are physically the most relevant. Numerical results fully
confirm the analytic expectations. The imaginary part of the
eigenvalues is always zero within the numerical accuracy. The
small-$c$ and large-$c$ behaviors are both in agreement with the
analytic predictions. As an additional code check we also verified
that the numerically computed eigenvalues satisfy the symmetry
relation (\ref{negs}).

\subsection{Large and pure-imaginary $c$ (prolate case)}
\label{prolate-large}

In the scalar case $s=0$, the asymptotic analysis of the eigenvalues
for large and pure-imaginary $c$ presents no particular difficulties,
and can be found in \cite{meixner,flammer}. A very accurate
approximation of the numerical eigenvalues in the left panel of
Fig.~\ref{fig:1}, which is valid within a few percent down to $c\simeq
2$, is:
\begin{eqnarray}\label{s0asy}
_{0}A_{lm}&=&(2L+1)|c_I|-(2L^2+2L+3-4m^2)2^{-2}
-\frac{1}{|c_I|}(2L+1)(L^2+L+3-8m^2)2^{-4} \\
&-&\frac{1}{|c_I|^2}[5(L^4+2L^3+7L+3)-48m^2(2L^2+2L+1)]2^{-6}+{}
 + {\cal O}(1/|c_I|^3)\,,\nonumber
\end{eqnarray}
where $L=l-|m|$ is the number of zeros of the scalar harmonics in the
interval $-1<x<1$. An expression accurate to ${\cal O}(1/|c_I|^6)$ can
be found in Eq.~(8.1.11) of \cite{flammer}.

Ref.~\cite{breuerbook} provides, to our knowledge, the first attempt
to generalize Eq.~(\ref{s0asy}) to general spin $s$. Unfortunately the
number of zeros of the SWSH given on page 115 of \cite{breuerbook} is
wrong, and was later corrected by Breuer {\it et al.} (Theorem 4.1 in
\cite{breuer}).  Therefore, although the derivation in
\cite{breuerbook} is essentially right, the expression for the leading
term is not.

Breuer's analysis for general spin was corrected in \cite{BCY}, but
two open issues remain. In the scalar case, the asymptotic solution
found in \cite{meixner,flammer} is a valid approximation to the
spheroidal harmonics only in a region far from the end-points $x=\pm
1$.  We expect the function, which is real, to have most zeros far
from the end-points; still, in principle we cannot discard the
possibility that the function does have a zero near the end-points.
As mentioned in the beginning of Sec.~\ref{subsec:4SWSHlarge,real},
the omission of a zero would lead to the wrong asymptotic expansion
for the eigenvalue (this is indeed what happened in \cite{breuer} in
the case of large, real $c$ and general spin $s$).

For general spin not only we run into the same problem, but we have the additional complication that the
solution is complex. However, in \cite{breuerbook,BCY} the number of zeros of the exact, {\it real} solution
for real $c$ was matched with the number of zeros of the real part of the asymptotic (complex) solution for
pure-imaginary $c$.  As indicated in \cite{meixner,flammer} for the case $s=0$, the number of zeros of the SWSH
is the same for $c^2\in (-\infty,+\infty)$, but this is only true for real functions. Numerical calculations
(see Sec.~\ref{EFnum} below, in particular Figs.~\ref{fig:mods2.ps} and \ref{fig:reims2.ps}) show that the
number of zeros of the real part and of the modulus of the SWSHs are both functions of $c$ (when $c$ is
pure-imaginary). Therefore it is not completely clear why the asymptotic analysis in \cite{breuerbook,BCY} for
the eigenvalue should be correct.

Below we provide analytic and numerical arguments in support of the
result obtained in \cite{breuerbook,BCY}.  We derive an asymptotic
solution for general spin-weight $s$ which is a valid approximation to
the SWSH near the end-points.  This ``outer'' solution does not
contain any zeros in the case $s=0$, and this explains why the
asymptotic behaviour of the eigenvalue is correct, at least for scalar
perturbations. We could not generalize this argument to
electromagnetic and gravitational perturbations, but we give numerical
evidence that, in all cases we considered, the asymptotic behaviour of
the eigenvalue given in \cite{BCY} for $|s|=1$ and $|s|=2$ is correct.

In order to find an asymptotic solution we define a new angular
wavefunction ${}_sy_{lm}(x)$ via
\begin{equation}
{}_sS_{lm}(x)=(1-x)^{k_+}(1+x)^{k_-}{}_sy_{lm}(x)\,,
\label{newwave}
\end{equation}
and change the independent variable setting $u\equiv\sqrt{2|c_I|}x$.
Substituting into Eq. (\ref{angularwaveeq}) we get
\begin{equation}
\begin{aligned}
\Bigg\{
&
\left(2|c_I|-u^2\right){d^2\over d u^2}-
2\left[\sqrt{2|c_I|}\left(k_+-k_-\right)+\left(k_++k_-+1\right)u\right]{d\over d u}
\\&
+{}_{s}A_{lm}+s(s+1)-\left(k_++k_-\right)\left(k_++k_-+1\right)-\frac{|c_I|u^2}{2}-
i\sqrt{2|c_I|}su
\Bigg\}{}_sy_{lm}=0
\end{aligned}
\end{equation}

When $|c_I| \rightarrow \infty$ and in the region $1/|c_I|\ll |x| \ll 1$,
the equation satisfied by the {\it real } part of ${}_sy_{lm}$ becomes
a parabolic cylinder equation.  Therefore the real part of the ``inner
solution'' ${}_sy_{lm}^{\text{inn}}$, which is a valid approximation
in this inner region, is a parabolic cylinder function $D_{L}$ with
$L$ zeros: $\Re \left({}_sy_{lm}^{\text{inn}}\right) (x)=
D_{L}\left(\sqrt{2|c_I|}x\right)$, where
\begin{equation}
_{s}A_{lm}=(2L+1)|c_I|+ {\cal O}(|c_I|^0) \,\,,\, |c_I| \rightarrow \infty \,.
\label{alm}
\end{equation}
This is essentially how the inner solution was found in
\cite{meixner,flammer,breuerbook,BCY}.  Note, however, that the
slightly different change of variable (\ref{newwave}) means that the
inner solution ${}_sS_{lm}^{\text{inn}}(x)$ we have obtained is always
regular at $x=\pm 1$, whereas the one given in \cite{breuerbook,BCY}
is not.

It is known \cite{AS} that the zeros of the parabolic cylinder
function $D_{L}\left(\sqrt{2|c_I|}x\right)$ lie within
$|x|<\sqrt{(2L+1)/|c_I|}$, so they are all within the inner region.
We can determine the value of $L$ by equating the number of zeros of
the parabolic cylinder function to the number of zeros of the {\it
real} part of the exact solution for {\it pure-imaginary} $c=\ii c_I$
(i.e., the real part of the prolate SWSH) in the region $1/|c_I|\ll
|x| \ll 1$.  Unfortunately, the latter number is not known. In
\cite{breuerbook,BCY} the number of zeros of the parabolic cylinder
function is instead equated to the number of zeros of the exact
solution for {\it real} $c$ (i.e., the oblate SWSH) in the whole
region $x\in (-1,+1)$, as given in \cite{breuer}. The value of $L$ so
determined turns out to be
\begin{equation}
L=l-{\rm max}(|m|,|s|).
\label{jdef}
\end{equation}
Higher order corrections in the asymptotic expansion can be obtained
as indicated in \cite{flammer}. We checked
Eqs.~(\ref{alm}) and (\ref{jdef}) numerically, solving Leaver's
angular continued fraction for $_{s}A_{lm}$ as a function of the
complex parameter $c$.

Now we take one more step towards understanding the asymptotic
behaviour given by Eqs.~(\ref{alm}) and (\ref{jdef}) providing an
``outer'' asymptotic solution, which is a valid approximation to the
SWSH in a region far from the origin.  A WKB-type expansion similar to
the one in \cite{CO} gives the result:

\begin{equation} \label{outer,prolateSWSH}
\begin{aligned}
{}_sS_{lm}^{\text{out}}(x)=
&
{}_sa_{lm} \left(1-x^2\right)^{-1/4} x^{L}
\left(1+\sqrt{1-x^2}\right)^{-L-1/2}\left(x-\ii\sqrt{1-x^2}\right)^{-s}e^{+|c_I|\sqrt{1-x^2}}+
\\ &
{}_sb_{lm} \left(1-x^2\right)^{-1/4} x^{-L-1}
\left(1+\sqrt{1-x^2}\right)^{L+1/2}\left(x-\ii\sqrt{1-x^2}\right)^{+s}e^{-|c_I|\sqrt{1-x^2}}
\end{aligned}
\end{equation}
where ${}_sa_{lm}$ and ${}_sb_{lm}$ are constants of integration.
This solution is valid when $|x|>>1/\sqrt{|c_I|}$ and $|x|\sim
1-O(|c_I|^\epsilon)$, with $-1<\epsilon\leq 0$.  Matching the outer and
inner solutions so that they coincide in their overlap region, we get
the following final expression for the outer solution (valid near
$x=\pm 1$):
\begin{equation}
{}_sS_{lm}^{\text{out},\pm 1}(x)=
(-\ii)^s 2^{L+1/2}\left(\pm\sqrt{2|c_I|}\right)^L e^{-3|c_I|/2} \left(1-x^2\right)^{-1/4} x^{L}
\left(1+\sqrt{1-x^2}\right)^{-L-1/2}\left(x-\ii\sqrt{1-x^2}\right)^{-s}e^{+|c_I|\sqrt{1-x^2}}\,.
\end{equation}

This solution formally diverges at $x=\pm 1$, and actually it is not
valid {\it at} the end-points, but only in a region {\it arbitrarily
close} to the end points. In such a region the exponential decay with
$|c_I|$ takes care of the diverging factor $\left(1-x^2\right)^{-1/4}$.
It can easily be checked that: (1) the outer solution has no zeros
when $s=0$; (2) its real part has no zeros within $x\in (-1,+1)$ when
$|s|=1$; (3) its real part has two zeros located at $x=\pm 1/\sqrt{2}$
when $|s|=2$.

We finally have a complete account of scalar perturbations ($s=0$).
The outer solution has no zeros, so all zeros of the asymptotic
solution are provided by the {\it inner} solution.  Equating the
number of zeros of the parabolic cylinder function to that of the
oblate SWSH is therefore valid in this case, which is why
Eqs.~(\ref{alm}) and (\ref{jdef}) are correct for $s=0$. For $s\neq 0$
we are one step closer to understanding the asymptotic behaviour. Now
we know all the zeros of the real part of the asymptotic solution, and
we ``only'' need to find out the number of zeros of the real part of
the SWSH for pure-imaginary frequency.  We have numerically checked in
many cases that Eqs. (\ref{alm}) and (\ref{jdef}) give the right
behaviour, and we believe them to be correct even for $s\neq 0$, but
only an analytic proof can settle the issue once and for all.

\begin{figure}[hbt]
\centerline{
\includegraphics[width=9cm,angle=0]{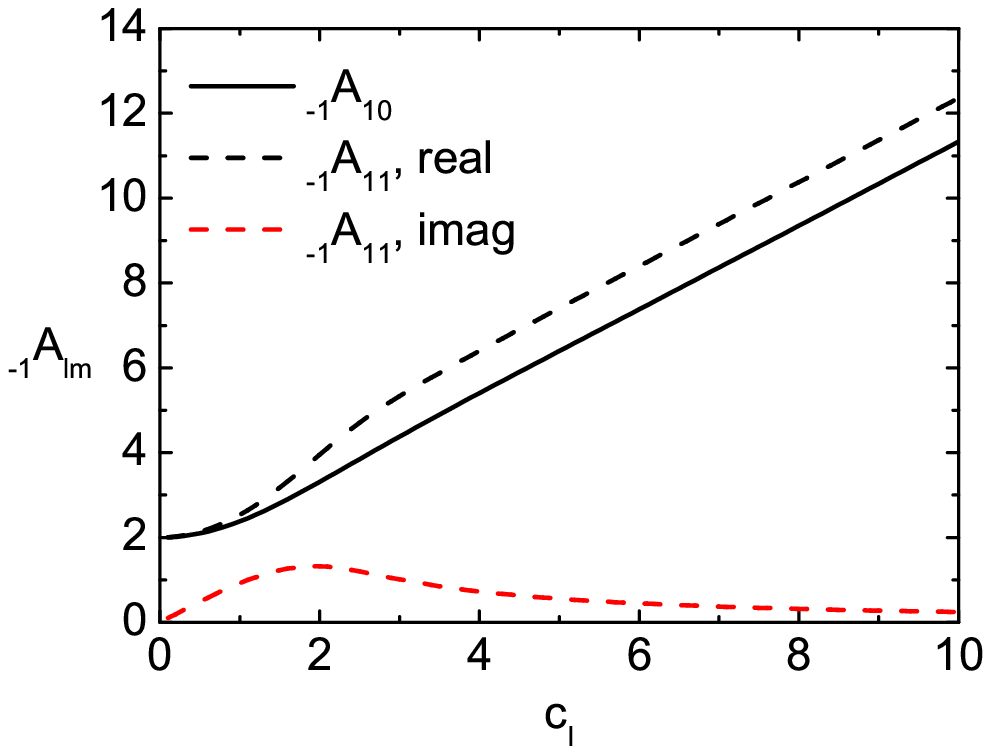}
\includegraphics[width=9cm,angle=0]{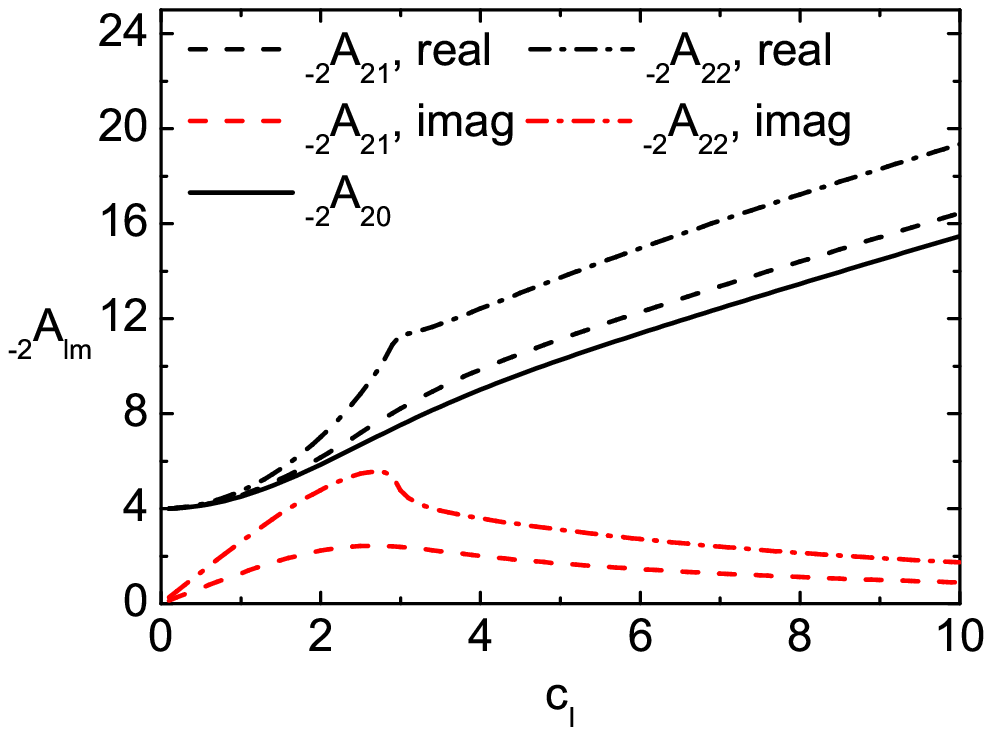}}
\caption{Prolate angular eigenvalues for $s=-1$ (left) and $s=-2$
(right). At variance with Fig.~\ref{fig:2}, the angular eigenvalue
${}_sA_{lm}$ is now complex. Lines limiting to
${}_sA_{lm}=l(l+1)-s(s+1)$ as $|c|\to 0$ are the real parts of the
eigenvalues. Lines approaching zero as $|c|\to 0$ are the imaginary
parts if $c_I<0$ (or their modulus if $c_I>0$: see property (iv) of
Sec.\ref{eqns}).}
\label{fig:2p}
\end{figure}

Fig.~\ref{fig:2p} shows prolate eigenvalues $_{s}A_{lm}$ for $s\neq 0$
and a few selected values of $(l,m)$, providing numerical support to
our analytic conclusions. The angular eigenvalue ${}_s A_{lm}$ is now
complex (unless $m=0$). We can still limit our calculations to
positive $m$'s because of the symmetry property (\ref{conj}). From
Fig.~\ref{fig:2p} we see that in the prolate case the real part
dominates as $c_I\to \infty$. The numerical results for any $s$ are
consistent with the linear growth predicted by Eqs.~(\ref{alm}) and
(\ref{jdef}).  A peculiar feature can be seen in the real part of
${}_{-2}A_{22}$, which seems to ``bend down'' at $c_I \simeq 3$. We
verified that the presence of this ``bending'' is not a numerical
artifact using both Leaver's method and the shooting method. There is
no discontinuity at the bending location, just a smooth change of
slope. Ref.~\cite{bertikerr2} found that, in the intermediate damping
regime, the real part of the Kerr quasinormal frequencies with $l=m=2$
tends to $\omega_R=2\Omega$, $\Omega$ being the angular velocity of
the horizon, unlike the frequency of modes with any other value of $l$
and $m$.  We suspect this anomalous behavior of quasinormal modes with
$l=m=2$ and $s=-2$ in the intermediate damping regime could be related
with the observed bending of the angular eigenvalue.

%

\begin{figure}[hbt]
\centerline{
\includegraphics[width=9cm,angle=0]{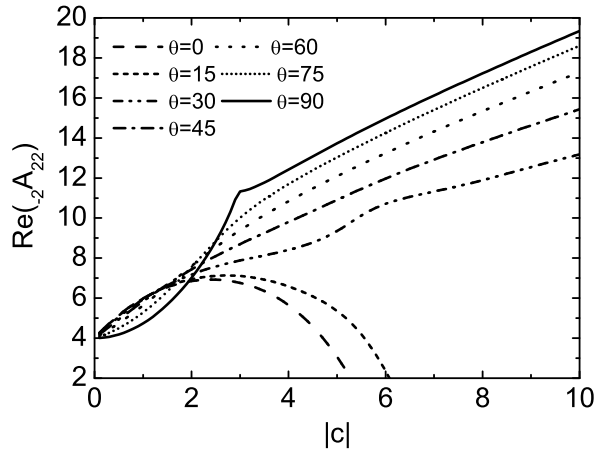}}
\caption{$\Re _{-2}A_{22}$ as a function of $|c|$ for different, fixed
values of the phase angle $\theta$.}
\label{fig:4}
\end{figure}

Finally, Fig.~\ref{fig:4} is a computational exercise to illustrate in
a specific case the transition from the prolate to the oblate regime.
We set $c=|c|e^{\ii\theta}$ and compute the real part of $_{-2}A_{22}$
along trajectories in the complex plane with fixed values of
$\theta\in[0,\pi/2]$. A phase angle $\theta=0$ corresponds to the
oblate case (Fig.~\ref{fig:2}) and a phase angle $\theta=\pi/2$ to the
prolate case (Fig.~\ref{fig:2p}). For intermediate values of $\theta$
the function $\Re _{-2}A_{22}(|c|)$ smoothly deforms from the prolate
to the oblate character. Notice also that the bending of the $l=m=2$
prolate eigenvalue we see at $|c|\simeq 3$ (see also
Fig.~\ref{fig:2p}) disappears immediately as $\theta$ becomes
nonzero. In our opinion, this provides some circumstantial evidence
that the bending is {\it not} due to a branch cut. We refer to
\cite{oguchi,barrowes} for an extensive discussion of branch cuts and
their effect on the prolate/oblate nature of the eigenvalues when
$s=0$.

%

\subsection{Numerical calculation of the prolate eigenfunctions}
\label{EFnum}

Once we have obtained the eigenvalues it is a simple matter to compute
the eigenfunctions for given values of $s$, $l$, $m$ and $c$: we just
need to compute the coefficients $a_p$ from the recursion relation
(\ref{recur})--(\ref{recur2}) and plug them into Leaver's series
solution, Eq.~(\ref{leavers}). The eigenfunctions for $s=0$ are well
known \cite{flammer}, and so are the eigenfunctions for oblate
harmonics (real $c$) with general spin $s$ \cite{CO}. Therefore here
we concentrate on {\it prolate} eigenfunctions with different values
of $c_I$. In black hole perturbation theory, according to Leaver's
convention on the Fourier transform \cite{leaver} stable perturbations
correspond to quasinormal frequencies with $\omega_I<0$. For this
reason, below we pick $c_I<0$. Using the symmetry property (iv) of
Sec.~\ref{eqns}, eigenvalues and eigenfunctions for $c_I>0$ are
trivially obtained by complex conjugation.

\begin{figure}[hbt]
\centerline{
\includegraphics[width=6cm,angle=0]{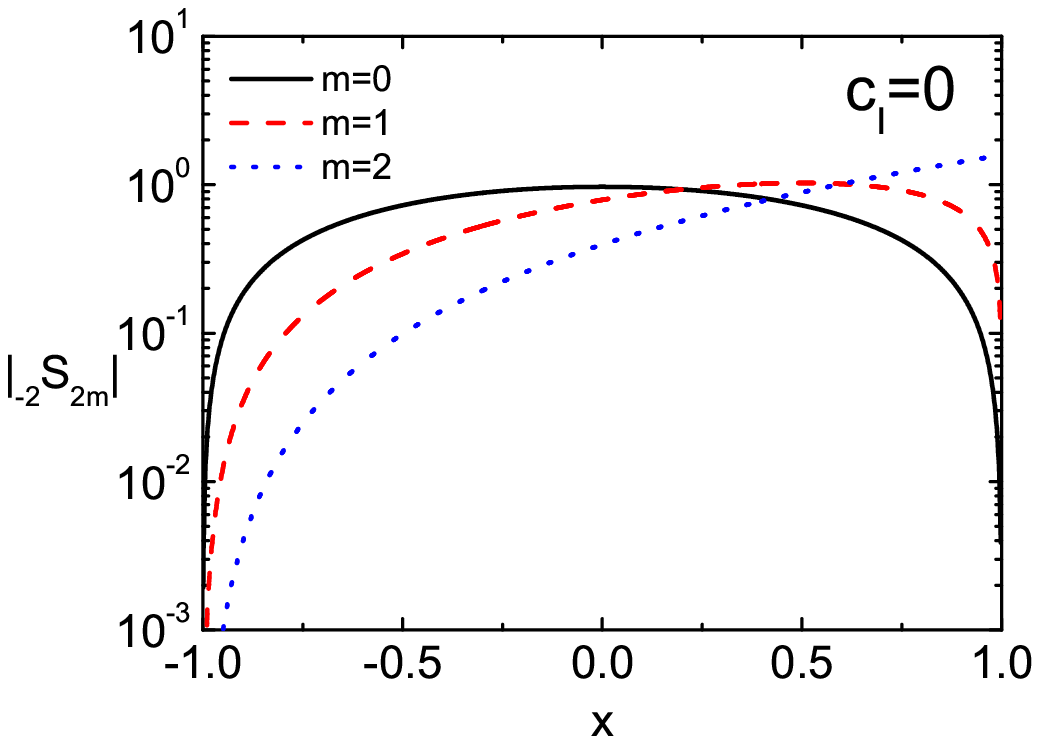}
\includegraphics[width=6cm,angle=0]{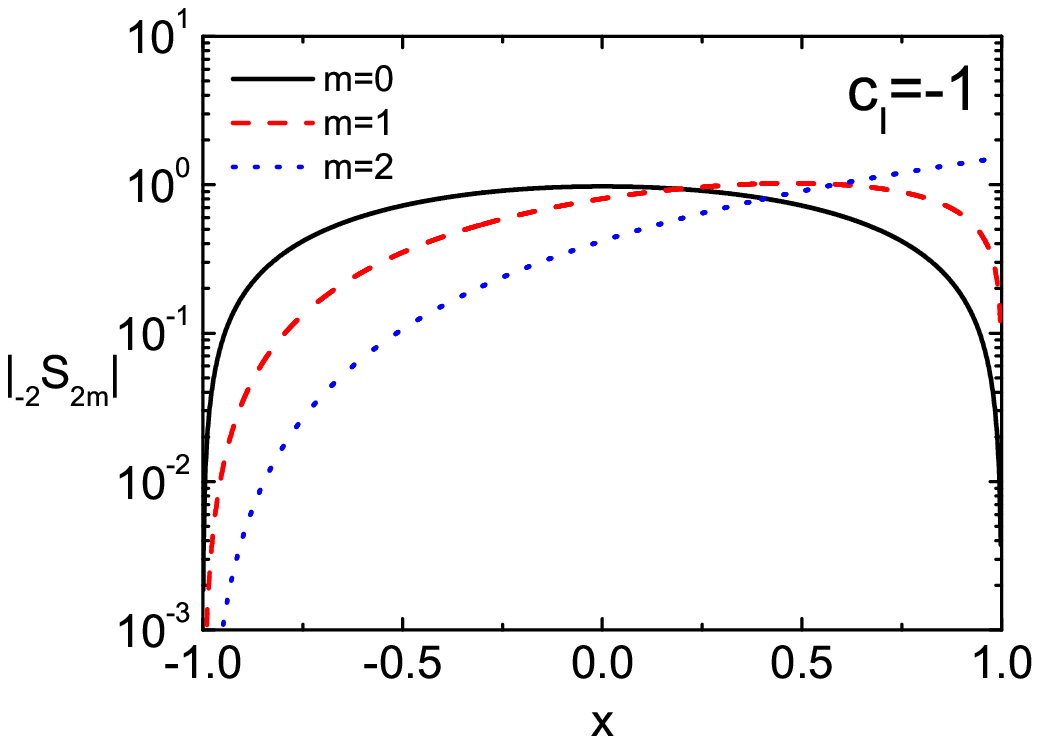}
\includegraphics[width=6cm,angle=0]{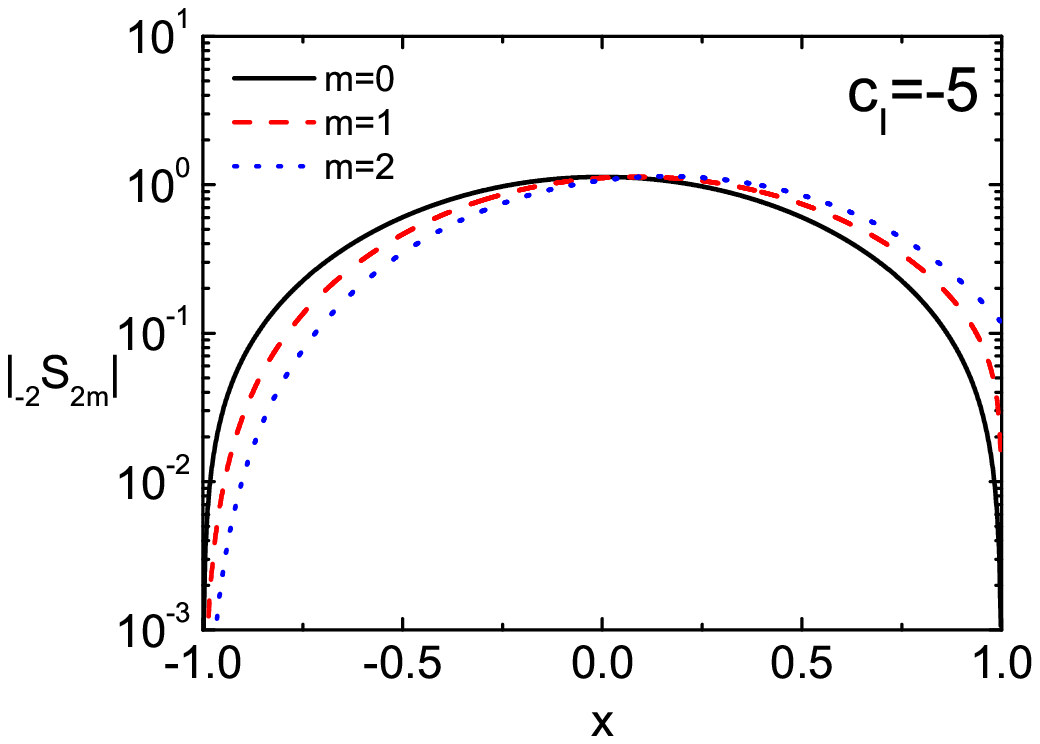}}
\centerline{
\includegraphics[width=6cm,angle=0]{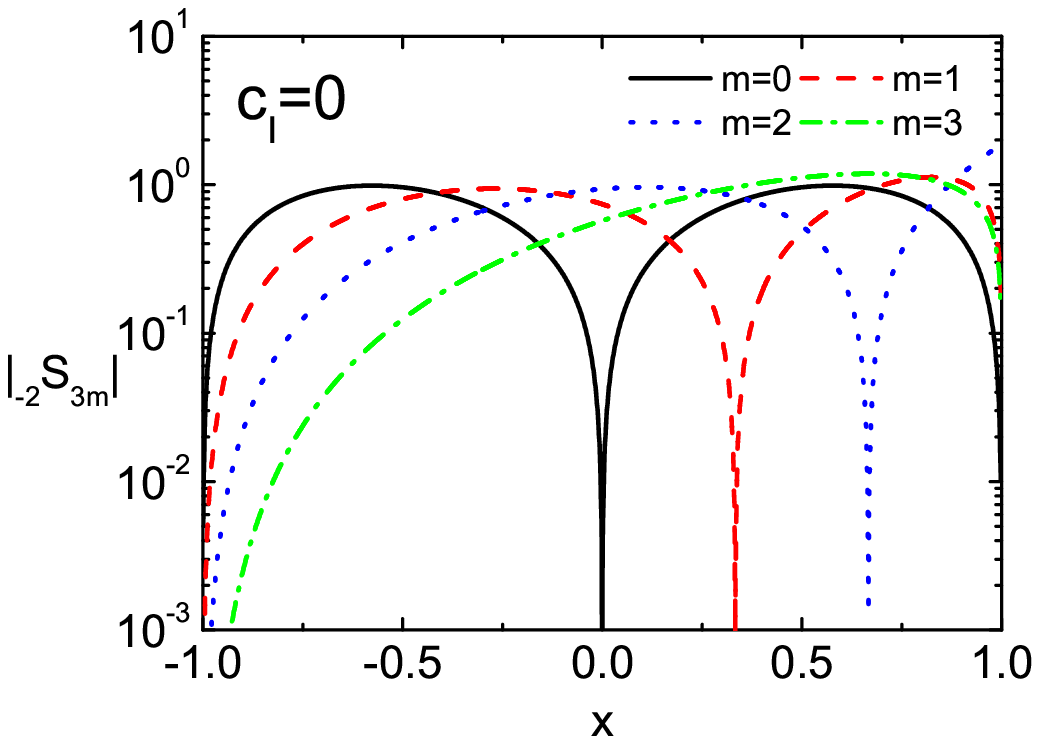}
\includegraphics[width=6cm,angle=0]{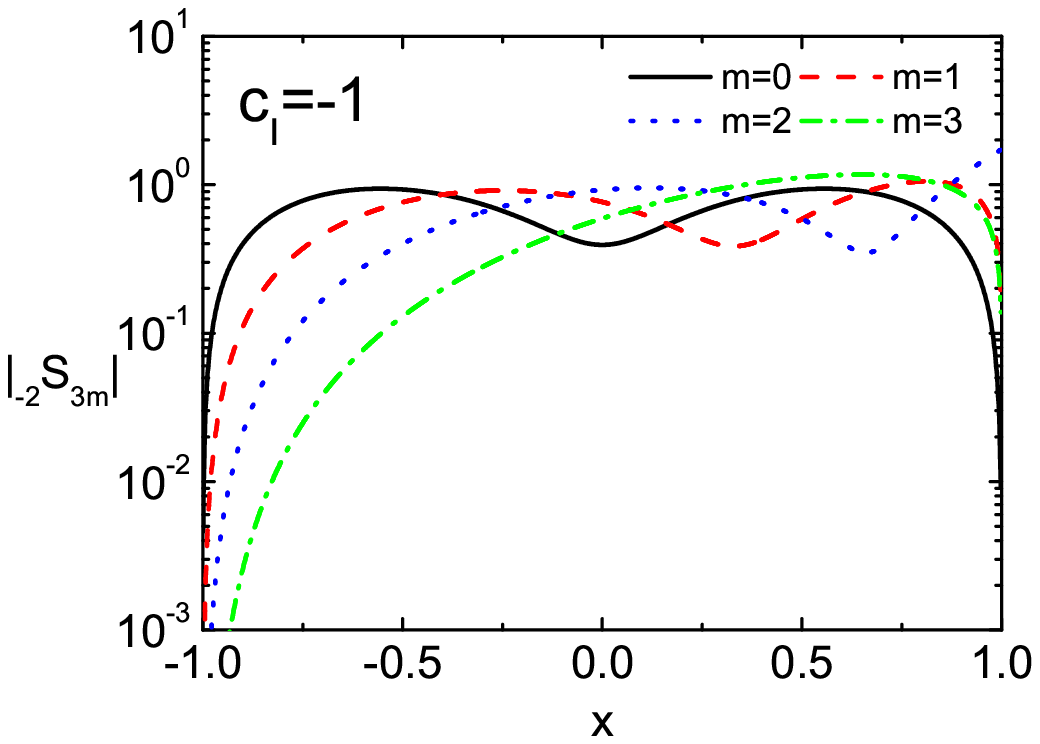}
\includegraphics[width=6cm,angle=0]{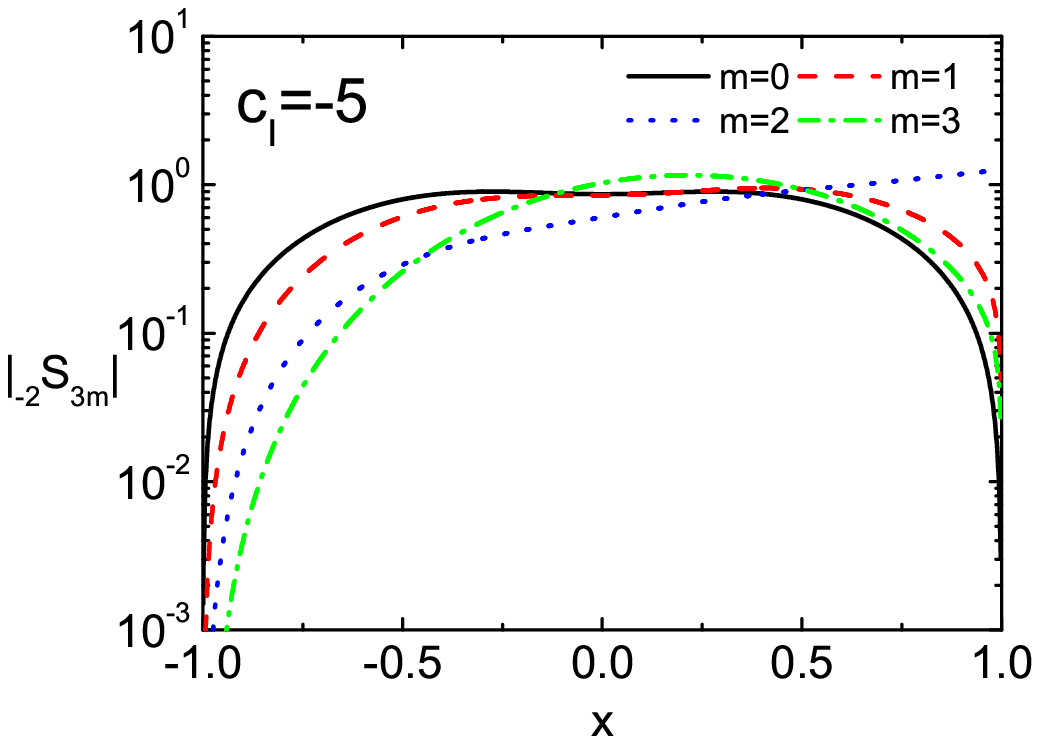}}
\caption{Modulus of the eigenfunction for prolate SWSHs with $s=-2$
and different values of $m$. The top row refers to $l=2$, the bottom
row to $l=3$. Left to right: $c_I=0,~-1$ and $-5$.
\label{fig:mods2.ps}}
\end{figure}

The modulus of prolate eigenfunctions with increasing values of
$|c_I|$ and $s=-2$ is plotted in Fig.~\ref{fig:mods2.ps}. The top row
refers to $l=2$, the bottom row to $l=3$. Each panel shows
eigenfunctions for different (positive) values of $m$. The left panels
refer to the simple case $c_I=0$. In this limit the eigenfunctions
reduce to spin-weighted ($s=-2$) {\it spherical} harmonics, which can
be thought of as being prolate and oblate at the same time
\cite{goldberg}. The ``spikes'' in this logarithmic plot are zeros of
the eigenfunctions, and their number is in agreement with
Eq.~(\ref{jdef}) for $c_I=0$. However, as anticipated,
Fig.~\ref{fig:mods2.ps} shows that (for $s\neq 0$) as soon as $c_I\neq
0$ the number of zeros in the region $-1<x<1$ changes.

\begin{figure}[hbt]
\centerline{
\includegraphics[width=9cm,angle=0]{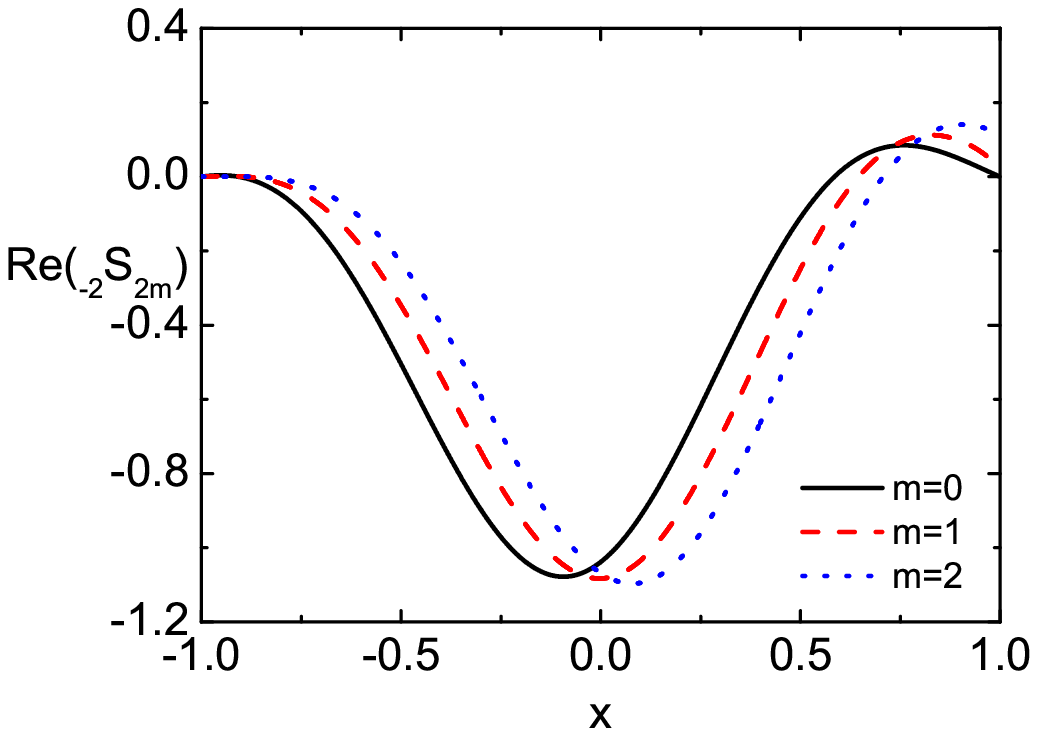}
\includegraphics[width=9cm,angle=0]{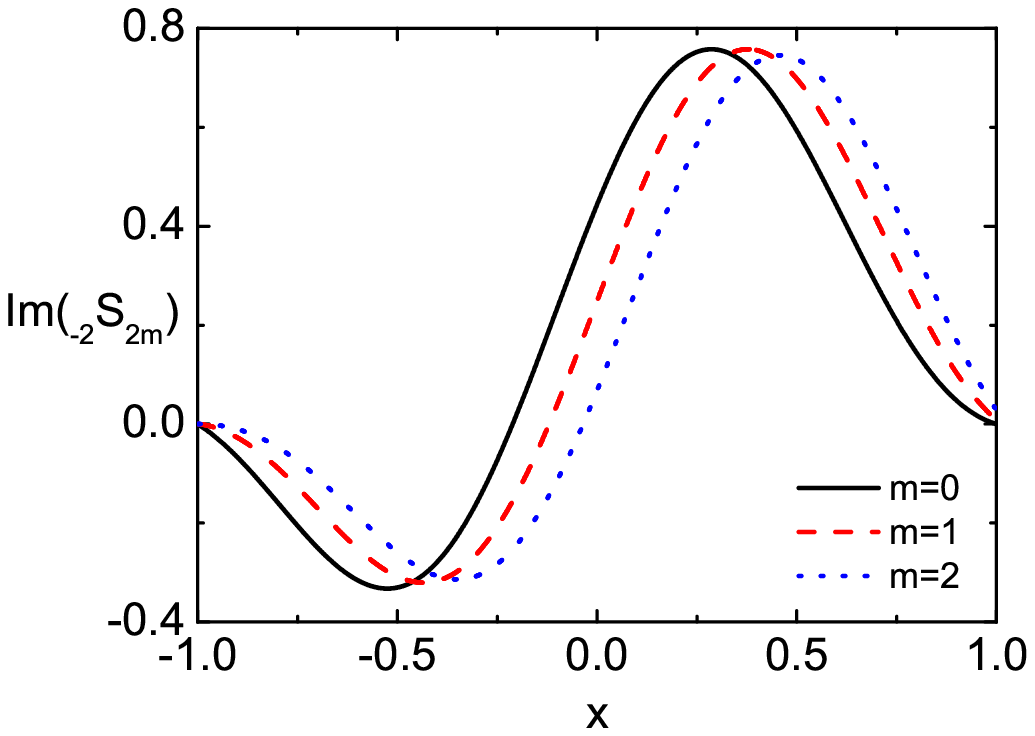}}
\caption{Real (left) and imaginary part (right) of prolate $l=2$,
$s=-2$ eigenfunctions for $c_I=-5$ and $m\geq 0$.
Note that the number of zeros of the real (or imaginary) part is not
zero, as predicted by Eq. (\ref{jdef}): the number of zeros of the
real part varies with $c_I$.
\label{fig:reims2.ps}}
\end{figure}

Fig.~\ref{fig:mods2.ps} only gives the modulus of the eigenfunctions,
to illustrate the point that prolate harmonics with spin-weight $s=-2$
have no zero for $-1<x<1$ when $c_I\neq 0$.  In
Fig.~\ref{fig:reims2.ps} we show {\it both} the real and imaginary
part of prolate eigenfunctions for $c_I=-5$.  Their modulus can be
read off the top right panel of Fig.~\ref{fig:mods2.ps}. The number of
zeros of the real part of the SWSH is not zero, as predicted by
Eq.~(\ref{jdef}) for $|m|\leq |s|=2$ and $l=2$. This illustrates that
the number of zeros of the real part of the eigenfunction (as well as
the number of zeros of its modulus: see Fig.~\ref{fig:mods2.ps})
depends on $c_I$.

\subsection{Scalar products of the eigenfunctions at the Kerr quasinormal
frequencies}
\label{KerrQNMs}

\begin{figure*}
\begin{center}
\centerline{
\includegraphics[width=9cm,angle=0]{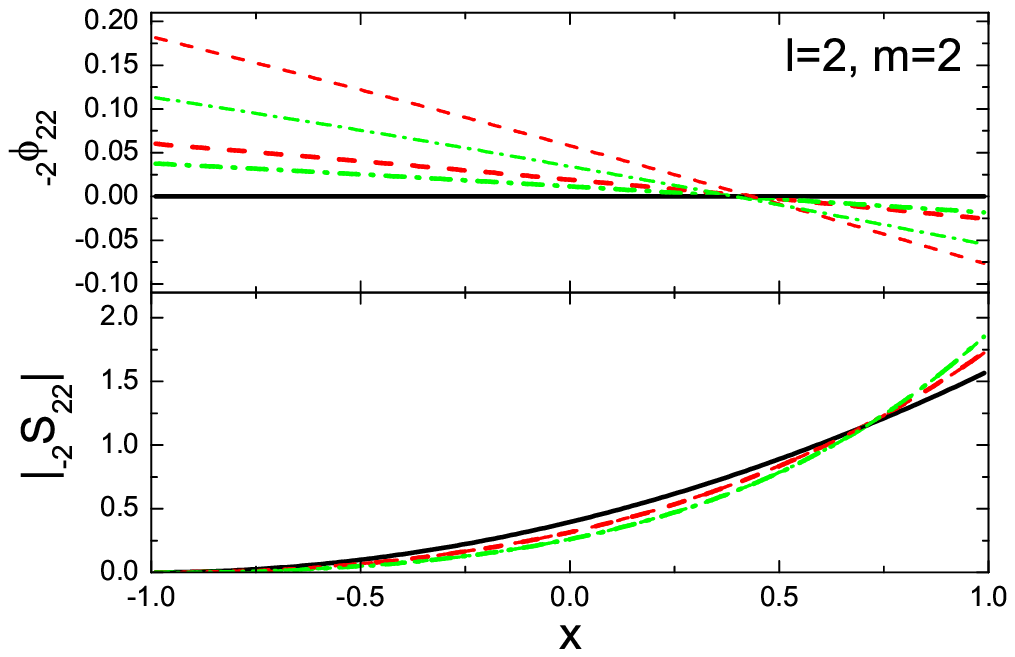}
\includegraphics[width=9cm,angle=0]{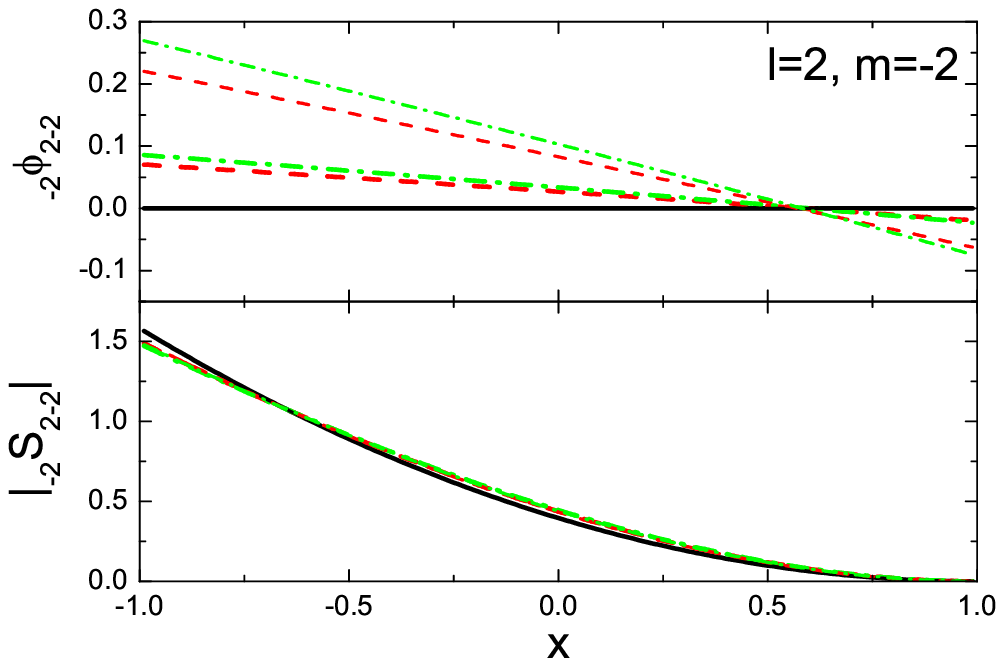}
}
\centerline{
\includegraphics[width=9cm,angle=0]{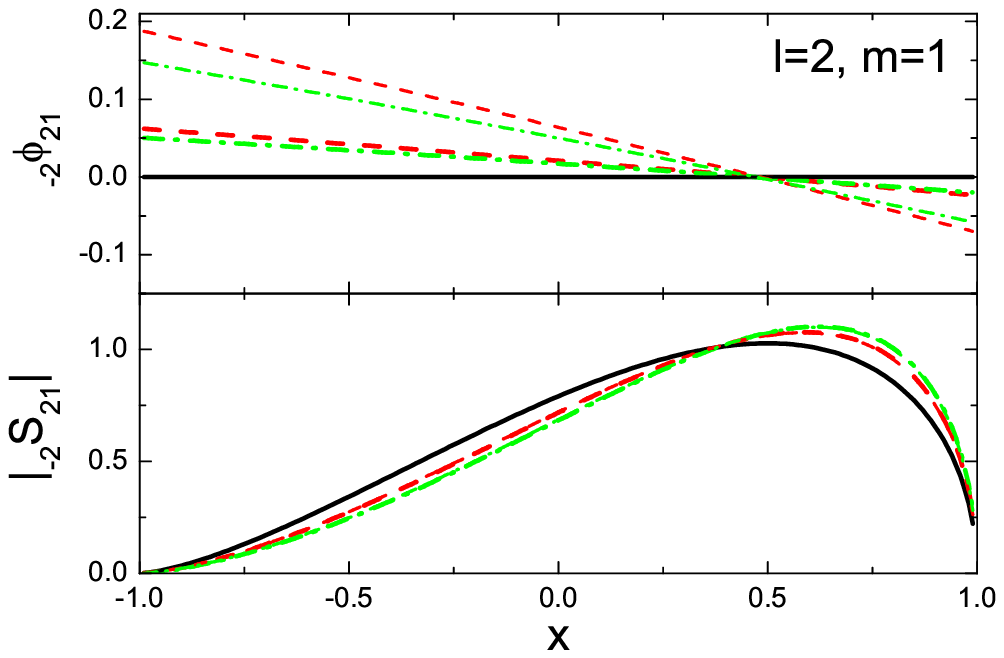}
\includegraphics[width=9cm,angle=0]{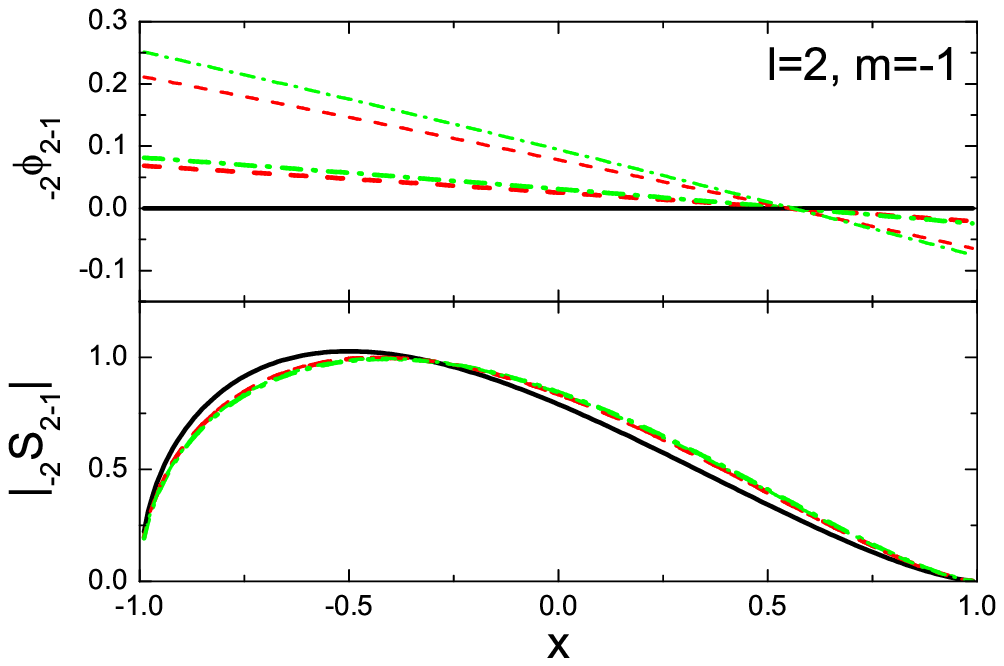}
}
\includegraphics[width=9cm,angle=0]{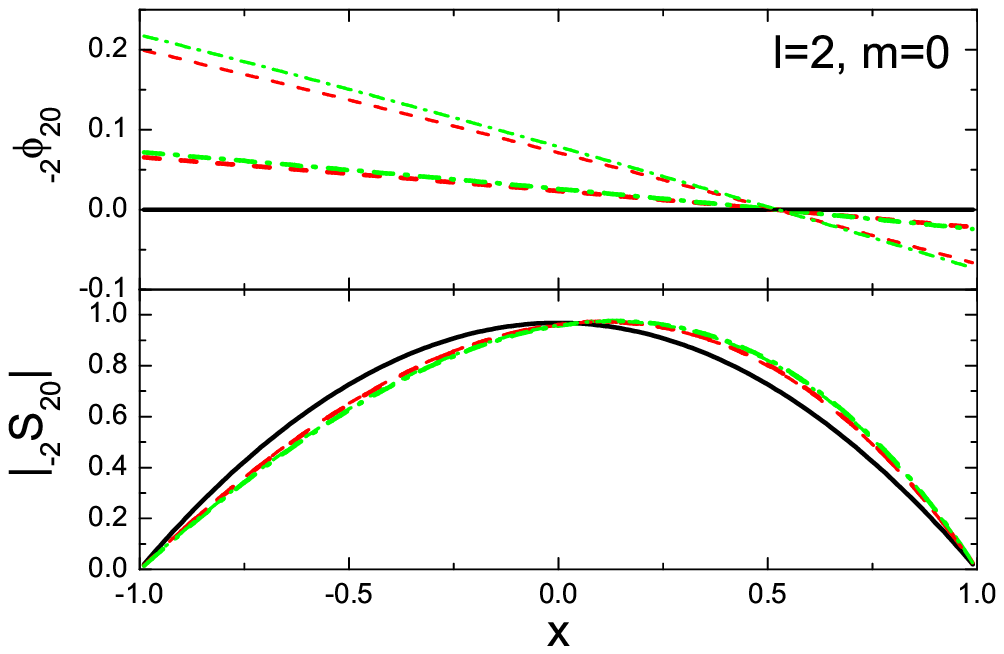}
\caption{Modulus $|_{s}S_{lm}|$ and phase $_{s}\phi_{lm}$ of the
angular eigenfunction for the fundamental mode (thick lines) and first
overtone (thin lines). The black hole's angular momentum $a/M=0$
(black, solid lines), $0.8$ (red, dashed lines) and $0.98$ (green,
dot-dashed lines). The phase (hence the imaginary part) usually grows
with the imaginary part of the mode frequencies, so it is larger for
higher overtones. The modulus of the eigenfunction gives an intuitive
picture of the angular pattern of the radiation: the radiation is
almost entirely concentrated around $x=\cos\theta=1$ (-1) for
corotating (counterrotating) modes, respectively. For axisymmetric
perturbations ($m=0$) most of the radiation is on the equatorial
plane, $x=\cos \theta=0$.
\label{fig:qnmef}}
\end{center}
\end{figure*}

As a useful application to general relativity of the numerical
algorithm described in Sec.~\ref{numerical}, here we study the angular
eigenfunctions corresponding to different quasinormal mode frequencies
and different values of $a/M$, for $l=2$. To our knowledge, these
eigenfunctions have never been shown before in the literature. The
procedure we use is by now standard and can be found in many papers
\cite{leaver,hisashi,bertikerr2,BCY,milan}, so here we only sketch the
basic idea.  In the Kerr space-time, linear gravitational
perturbations are described by a pair of coupled differential
equations: one for the angular part of the perturbations, and the
other for the radial part.  The radial equation resulting from the
separation of the field equations in the Kerr background is given
(eg.) in \cite{teukolsky,leaver}. The angular equation is a SWSH
equation with $c=a\omega$. Boundary conditions for the two equations
can be cast as a couple of three-term continued fraction relations of
the form (\ref{cfeq}). Computing quasinormal frequencies $\omega$ and
the related angular eigenvalues is simple. For assigned values of
$s,l,m,a$ and $\omega$, first find the angular separation constant
$_{s}A_{lm}$ looking for zeros of the {\it angular} continued
fraction. Then replace the corresponding eigenvalue into the {\it
radial} continued fraction, and look for its zeros as a function of
$\omega$.

Once we know the quasinormal mode frequencies and the angular
eigenvalues, we can find the angular eigenfunction following the
procedure of Sec.~\ref{numerical}.  The results are plotted in
Fig.~\ref{fig:qnmef} for different gravitational ($s=-2$) quasinormal
frequencies with $l=2$, $|m|\leq l$ and different values of $a/M$.
The modulus of the eigenfunction is scarcely affected by rotation and
almost unaffected by the overtone index, at least for slowly damped
modes. The phase (hence the relative magnitude of the imaginary part
of the eigenfunctions) is more sensitive to angular momentum and mode
damping. It is generally larger for counterrotating modes, and it
grows with the overtone index $N$.  In general, the magnitude of the
phase seems to be related with the magnitude of the imaginary part of
the quasinormal frequencies. For corotating modes in the extremal
limit the imaginary part of quasinormal frequencies goes to zero, as
first shown by Detweiler \cite{detweiler}. This explains what, at
first sight, might look like a puzzling feature of the plots: for
near-extremal Kerr black holes ($a/M=0.98$, green lines) the phase is
actually {\it smaller} than for black holes with $a/M=0.80$ if the
modes are corotating. The opposite is true for counterrotating modes,
whose imaginary part does not tend to zero in the extremal limit.

In analysing data from gravitational waveforms emitted by an
oscillating black hole it is necessary to compute the scalar products
\cite{bertiringdown}
\be \int~{_{-2}S_{lm}}^*(a\omega_{lmN}) ~
_{-2}S_{l'm'}(a\omega_{l'm'N'}) d\Omega=\alpha_{mll'NN'}(a)
\delta_{m,m'}\,. \label{ccoeff}
\ee
Here and in the following we append to the quasinormal frequencies the
angular indices $(l,~m)$ and a third index (the ``overtone'' index
$N=0,~1,~2\dots$) that sorts frequencies by the magnitude of their
imaginary part: in the limit $a/M=0$, modes with larger $N$ have
larger imaginary part and damp faster. The angular momentum-dependent
quantity $\alpha_{mll'NN'}(a)$ will be determined below, and the
Kronecker symbol $\delta_{m,m'}$ comes from the $e^{\ii m
\phi}$-dependence of the harmonics. For $a/M=0$ the spin-weighted
spheroidal harmonics reduce to spin-weighted spherical harmonics
${}_sY_{lm}$ \cite{goldberg}, for which
\be
\int~{_{-2}Y_{lm}}^*~ _{-2}Y_{l'm'} d\Omega=\delta_{l,l'} \delta_{m,m'}\,.
\ee
We evaluated SWSHs and their scalar products by two independent,
approximate calculations.

In the first calculation we use the sixth-order polynomial
approximation for the separation constant $_{s}A_{lm}$ computed by
Press and Teukolsky (Table I in Ref.~\cite{PT}). We plug this
expansion in the spheroidal wave equation and integrate the equation
numerically, imposing the appropriate boundary conditions at one
extremum of the integration interval. In principle we should expect
the agreement with Leaver's method to be good only for small values of
$a\omega$. Surprisingly, results for the eigenfunctions and the scalar
products turn out to be very close also for large $a\omega$, where the
polynomial approximation of the ``true'' separation constant is not so
accurate. Even for $a/M=0.98$ the approximate scalar products agree
with Leaver's method within $\sim 5\%$.

As a second independent, analytic check we solved the spheroidal wave
equation expanding in powers of $a\omega$ and using standard
perturbation theory. For $a\omega=0$ the solutions are the
spin-weighted spherical harmonics \cite{goldberg}. The next order
correction was found by Press and Teukolsky (Eq. (3.7) of
Ref.~\cite{PT}). Using their approximate solution we can show that, to
leading order,
\be \label{swshsp}
\alpha_{mll'NN'}(a)\simeq 4a\left[ \omega_{lmN}^* \theta(l,l',m)
+\omega_{l'm'N'}\theta(l',l,m) \right]\,,\,\,\,l \neq l'\,,
\ee
where $\alpha_{mll'NN'}(a)$ has been defined in Eq.~(\ref{ccoeff}),
\be
\theta(l,l',m)\equiv \frac{\sqrt{2l+1}}{\sqrt{2l'+1}[l(l+1)-l'(l'+1)]}
\langle l1m0|l'm\rangle \langle l120|l'2\rangle\,, \ee
and $\langle j_1j_2m_1m_2|JM\rangle$ is a Clebsch-Gordan
coefficient. For $l=l'$ the same first-order perturbative analysis
gives $\alpha_{mll'nn'}(a)\simeq 1$.

In Table \ref{tab:pressanalytical} we compare values of the scalar
products obtained using our two approximate methods with the results of
Leaver's method. There we focus on the worst possible scenario of a
nearly maximally rotating Kerr black hole ($a/M=0.98$). Both
approximations work really well, providing a powerful consistency
check of the numerical calculation.

\begin{table}[hbt]
\centering \caption{\label{tab:pressanalytical} Comparison between
different methods to evaluate the scalar product of spheroidal
harmonics for a nearly-maximally rotating Kerr black hole
($a/M=0.98$). We computed all numbers for $N=N'=0$, so we only list
the corresponding values of $l$,~$l'$ and~$m$. The most accurate
result comes from Leaver's method, but an expansion of the separation
constant in powers of $a\omega$, followed by a direct integration of
the angular equation, is accurate to within a few percent.  The first
order analytic expansion~(\ref{swshsp}) also gives a surprisingly good
approximation to the numerical results.}
\begin{tabular}{||c||c|c|c||}  \hline
\multicolumn{4}{||c||}{{\rm Scalar product}}\\
\hline
$(l,l',m)$ &{\rm Leaver}& {\rm Polynomial}&{\rm Analytic}\\
\hline
$(2,3,2)$  &$0.0275-0.0144\ii$ &$0.0272-0.0150\ii$ &$0.0279-0.0154\ii$\\
$(2,3,1)$  &$0.0525-0.0258\ii$ &$0.0517-0.0273\ii$ &$0.0510-0.0269\ii$\\
$(2,3,0)$  &$0.0608-0.0361\ii$ &$0.0598-0.0379\ii$ &$0.0591-0.0375\ii$\\
$(2,3,-1)$  &$0.0547-0.0378\ii$ &$0.0539-0.0391\ii$ &$0.0543-0.0394\ii$\\
$(2,3,-2)$  &$0.0398-0.0308\ii$ &$0.0394-0.0314\ii$ &$0.0409-0.0327\ii$\\
\hline
$(3,4,2)$  &$0.0393-0.0204\ii$ &$0.0389-0.0212\ii$ &$0.0345-0.0187\ii$\\
$(3,4,1)$  &$0.0487-0.0289\ii$ &$0.0482-0.0298\ii$ &$0.0452-0.0279\ii$\\
$(3,4,0)$  &$0.0487-0.0335\ii$ &$0.0484-0.0342\ii$ &$0.0480-0.0338\ii$\\
$(3,4,-1)$  &$0.0440-0.0338\ii$ &$0.0438-0.0341\ii$ &$0.0457-0.0355\ii$\\
$(3,4,-2)$  &$0.0363-0.0303\ii$ &$0.0363-0.0304\ii$ &$0.0395-0.0331\ii$\\
\hline
\end{tabular}
\end{table}

For reference, the constants $\alpha_{mll'nn'}(a)$ computed using
Leaver's method are given in Tables~\ref{tab:a040} and \ref{tab:a049}
for fixed values of $m=m'$ (for $m\neq m'$ the scalar product is
zero).  In the context of gravitational wave data analysis, it is most
useful to consider scalar products between the quasinormal modes which
are more likely to be excited during collapse or merger: say, the
first three Kerr overtones $(N,~N')=0,~1,~2$ with $(l,~l')=2,~3,~4$
and $|m|\leq 2$. Tables~\ref{tab:a040} and \ref{tab:a049} show that,
to a reasonable level of approximation, we can assume
$\alpha_{mll'nn'}(a)\simeq \delta_{ll'}$, that is,
\be\label{Slm-orth}
\int~{_{-2}S_{lm}}^*(a\omega_{lmN}) _{-2}S_{l'm'}(a\omega_{l'm'N'})
d\Omega\simeq \delta_{l,l'} \delta_{m,m'}\,.
\ee
The errors on this approximate relation increase with the black hole's
angular momentum $a/M$. They are typically larger for small $m$, but in
most cases the approximate formula is valid to within $\sim 10 \%$, at
least for cases of physical interest (say, for $l<5$).

\begin{table}[hbt]
\centering
\caption{\label{tab:a040} Scalar products of the spin weighted
spheroidal harmonics for $a/M=0.80$. To save space, we omit the
leading zeros and the $\ii$'s in the imaginary parts (so, for example,
$.9994\pm.0255$ actually means $0.9994\pm0.0255\ii$). Entries
replaced by an asterisk can be obtained from the symmetric entries by
complex conjugation: i.e.,
$\alpha_{mll'NN'}(a)=\alpha^*_{ml'lN'N}(a)$.}
\begin{tabular}{|c|c|ccc|ccc|ccc|}
\hline
\multicolumn{2}{|c|}{$m$}&\multicolumn{9}{|c|}{$a/M=0.80$}\\
\hline
\multicolumn{2}{|c|}{$2$}
&\multicolumn{3}{|c|}{$l'=2$}
&\multicolumn{3}{|c|}{$l'=3$}
&\multicolumn{3}{|c|}{$l'=4$}\\
\hline
$l$&$N$&$N'=0$ &$N'=1$ &$N'=2$ &$N'=0$ &$N'=1$ &$N'=2$ &$N'=0$ &$N'=1$ &$N'=2$\\
\hline
   &$0$&1            &.9994-.0255&.9976-.0510&.0339-.0221&.0359-.0443&.0392-.0665&-.0016-.0005&-.0024-.0007&-.0036-.0011\\
$2$&$1$&*&1            &.9994-.0255&.0370-.0441&.0405-.0661&.0454-.0880&-.0026-.0013&-.0039-.0018&-.0057-.0024\\
   &$2$&*&*&1            &.0422-.0665&.0473-.0881&.0538-.1096&-.0044-.0024&-.0062-.0032&-.0085-.0043\\
\hline
   &$0$&*&*&*&1&.9984+.0432&.9937+.0869&.0357-.0260&.0389-.0503&.0449-.0743\\
$3$&$1$&*&*&*&*&1            &.9984+.0439&.0352-.0535&.0394-.0778&.0465-.1016\\
   &$2$&*&*&*&*&*            &1            &.0335-.0811&.0388-.1054&.0469-.1291\\
\hline
   &$0$&*&*&*&*&*&*&1&.9960+.0809&.9842+.1614\\
$4$&$1$&*&*&*&*&*&*&*&1            &.9960+.0813\\
   &$2$&*&*&*&*&*&*&*&*            &1            \\
\hline
\multicolumn{2}{|c|}{$1$}
&\multicolumn{3}{|c|}{$l'=2$}
&\multicolumn{3}{|c|}{$l'=3$}
&\multicolumn{3}{|c|}{$l'=4$}\\
\hline
$l$&$N$&$N'=0$ &$N'=1$ &$N'=2$ &$N'=0$ &$N'=1$ &$N'=2$ &$N'=0$ &$N'=1$ &$N'=2$\\
\hline
   &$0$&1&.9995+.0074&.9980+.0158&.0455-.0295&.0485-.0581&.0540-.0868&-.0023-.0007&-.0033-.0010&-.0049-.0015\\
$2$&$1$&*&1            &.9995+.0083&.0484-.0600&.0533-.0883&.0607-.1165&-.0038-.0020&-.0056-.0026&-.0079-.0035\\
   &$2$&*&*            &1            &.0528-.0913&.0596-.1193&.0688-.1468&-.0066-.0038&-.0091-.0048&-.0121-.0062\\
\hline
   &$0$&*&*&*&1&.9968+.0677&.9869+.1356&.0390-.0291&.0432-.0554&.0512-.0811\\
$3$&$1$&*&*&*&*&1            &.9967+.0686&.0372-.0608&.0424-.0871&.0514-.1127\\
   &$2$&*&*&*&*&*            &1            &.0328-.0927&.0390-.1191&.0491-.1448\\
\hline
   &$0$&*&*&*&*&*&*&1&.9944+.0978&.9776+.1949\\
$4$&$1$&*&*&*&*&*&*&*&1            &.9943+.0985\\
   &$2$&*&*&*&*&*&*&*&*            &1            \\
\hline
\multicolumn{2}{|c|}{$0$}
&\multicolumn{3}{|c|}{$l'=2$}
&\multicolumn{3}{|c|}{$l'=3$}
&\multicolumn{3}{|c|}{$l'=4$}\\
\hline
$l$&$N$&$N'=0$ &$N'=1$ &$N'=2$ &$N'=0$ &$N'=1$ &$N'=2$ &$N'=0$ &$N'=1$ &$N'=2$\\
\hline
   &$0$&1&.9985+.0420&.9937+.0870&.0484-.0329&.0524-.0635&.0603-.0940&-.0025-.0007&-.0035-.0009&-.0049-.0013\\
$2$&$1$&*&1            &.9983+.0451&.0504-.0690&.0563-.0992&.0661-.1291&-.0045-.0022&-.0062-.0028&-.0085-.0037\\
   &$2$&*&*            &1            &.0520-.1074&.0599-.1372&.0717-.1665&-.0082-.0044&-.0109-.0055&-.0139-.0070\\
\hline
   &$0$&*&*&*&1&.9946+.0924&.9782+.1853&.0386-.0300&.0436-.0562&.0535-.0817\\
$3$&$1$&*&*&*&*&1            &.9944+.0942&.0353-.0639&.0413-.0902&.0522-.1158\\
   &$2$&*&*&*&*&*            &1            &.0276-.0981&.0346-.1247&.0465-.1506\\
\hline
   &$0$&*&*&*&*&*&*&1&.9927+.1141&.9704+.2275\\
$4$&$1$&*&*&*&*&*&*&*&1            &.9925+.1154\\
   &$2$&*&*&*&*&*&*&*&*            &1            \\
\hline
\multicolumn{2}{|c|}{$-1$}
&\multicolumn{3}{|c|}{$l'=2$}
&\multicolumn{3}{|c|}{$l'=3$}
&\multicolumn{3}{|c|}{$l'=4$}\\
\hline
$l$&$N$&$N'=0$ &$N'=1$ &$N'=2$ &$N'=0$ &$N'=1$ &$N'=2$ &$N'=0$ &$N'=1$ &$N'=2$\\
\hline
   &$0$&1&.9963+.0789&.9834+.1658&.0439-.0321&.0488-.0606&.0590-.0892&-.0023-.0005&-.0030-.0005&-.0040-.0007\\
$2$&$1$&*&1            &.9954+.0877&.0444-.0695&.0511-.0977&.0630-.1257&-.0043-.0020&-.0058-.0023&-.0075-.0031\\
   &$2$&*&*            &1            &.0410-.1117&.0497-.1397&.0637-.1673&-.0087-.0040&-.0110-.0050&-.0134-.0065\\
\hline
   &$0$&*&*&*&1&.9922+.1166&.9676+.2350&.0354-.0289&.0410-.0533&.0522-.0769\\
$3$&$1$&*&*&*&*&1            &.9916+.1208&.0308-.0625&.0373-.0872&.0495-.1112\\
   &$2$&*&*&*&*&*            &1            &.0199-.0966&.0274-.1220&.0406-.1468\\
\hline
   &$0$&*&*&*&*&*&*&1&.9908+.1294&.9628+.2586\\
$4$&$1$&*&*&*&*&*&*&*&1            &.9905+.1320\\
   &$2$&*&*&*&*&*&*&*&*            &1            \\
\hline
\multicolumn{2}{|c|}{$-2$}
&\multicolumn{3}{|c|}{$l'=2$}
&\multicolumn{3}{|c|}{$l'=3$}
&\multicolumn{3}{|c|}{$l'=4$}\\
\hline
$l$&$N$&$N'=0$ &$N'=1$ &$N'=2$ &$N'=0$ &$N'=1$ &$N'=2$ &$N'=0$ &$N'=1$ &$N'=2$\\
\hline
   &$0$&1&.9927+.1170&.9656+.2523&.0327-.0257&.0376-.0476&.0486-.0700&-.0017-.0002&-.0020-.0001&-.0025-.0002\\
$2$&$1$&*&1            &.9899+.1377&.0319-.0577&.0382-.0794&.0505-.1016&-.0032-.0013&-.0041-.0015&-.0050-.0020\\
   &$2$&*&*            &1            &.0237-.0963&.0319-.1183&.0461-.1408&-.0071-.0027&-.0086-.0034&-.0101-.0046\\
\hline
   &$0$&*&*&*&1&.9895+.1397&.9556+.2841&.0298-.0255&.0354-.0465&.0469-.0668\\
$3$&$1$&*&*&*&*&1            &.9882+.1479&.0245-.0560&.0310-.0775&.0434-.0984\\
   &$2$&*&*&*&*&*            &1            &.0118-.0872&.0190-.1097&.0324-.1318\\
\hline
   &$0$&*&*&*&*&*&*&1&.9890+.1436&.9551+.2879\\
$4$&$1$&*&*&*&*&*&*&*&1            &.9884+.1478\\
   &$2$&*&*&*&*&*&*&*&*            &1            \\
\hline
\end{tabular}
\end{table}

\begin{table}[hbt]
\centering
\caption{\label{tab:a049} Scalar products of the spin weighted
spheroidal harmonics for $a/M=.98$. To save space, we omit the
leading zeros and the $\ii$'s in the imaginary parts (so, for example,
$.9997\pm.0213$ actually means $0.9997\pm0.0213\ii$). Entries
replaced by an asterisk can be obtained from the symmetric entries by
complex conjugation: i.e.,
$\alpha_{mll'NN'}(a)=\alpha^*_{ml'lN'N}(a)$.}
\begin{tabular}{|c|c|ccc|ccc|ccc|}
\hline
\multicolumn{2}{|c|}{$m$}&\multicolumn{9}{|c|}{$a/M=.98$}\\
\hline
\multicolumn{2}{|c|}{$2$}
&\multicolumn{3}{|c|}{$l'=2$}
&\multicolumn{3}{|c|}{$l'=3$}
&\multicolumn{3}{|c|}{$l'=4$}\\
\hline
$l$&$N$&$N'=0$ &$N'=1$ &$N'=2$ &$N'=0$ &$N'=1$ &$N'=2$ &$N'=0$ &$N'=1$ &$N'=2$\\
\hline
   &$0$&1                    &.9997-.0213&.9987-.0427&.0275-.0144&.0277-.0294&.0280-.0437&-.0027+.0000&-.0033+.0001&-.0041+.0004\\
$2$&$1$&*&1                  &.9997-.0214&.0284-.0277&.0293-.0428&.0303-.0570&-.0032-.0001&-.0040-.0000&-.0052+.0001\\
   &$2$&*&*&1                &.0299-.0411&.0315-.0560&.0331-.0702&-.0039-.0003&-.0051-.0003&-.0066-.0002\\
\hline
   &$0$&*&*&*&1              &.9993+.0269&.9975+.0524&.0393-.0204&.0399-.0412&.0402-.0613\\
$3$&$1$&*&*&*&*&1            &.9994+.0256&.0400-.0399&.0413-.0605&.0422-.0804\\
   &$2$&*&*&*&*&*&1          &.0407-.0583&.0425-.0787&.0440-.0984\\
\hline
   &$0$&*&*&*&*&*&*&1        &.9973+.0662&.9897+.1292\\
$4$&$1$&*&*&*&*&*&*&*&1      &.9975+.0633\\
   &$2$&*&*&*&*&*&*&*&*&1    \\
\hline
\multicolumn{2}{|c|}{$1$}
&\multicolumn{3}{|c|}{$l'=2$}
&\multicolumn{3}{|c|}{$l'=3$}
&\multicolumn{3}{|c|}{$l'=4$}\\
\hline
$l$&$N$&$N'=0$ &$N'=1$ &$N'=2$ &$N'=0$ &$N'=1$ &$N'=2$ &$N'=0$ &$N'=1$ &$N'=2$\\
\hline
   &$0$&1            &.9997+.0031&.9990+.0063&.0525-.0258&.0532-.0528&.0533-.0802&-.0036-.0006&-.0047-.0009&-.0064-.0013\\
$2$&$1$&*&1            &.9998+.0031&.0569-.0492&.0590-.0758&.0605-.1028&-.0046-.0017&-.0062-.0023&-.0085-.0029\\
   &$2$&*&*&1            &.0638-.0685&.0669-.0947&.0695-.1210&-.0057-.0031&-.0078-.0040&-.0105-.0050\\
\hline
   &$0$&*&*&*&1&.9971+.0632&.9886+.1255&.0487-.0289&.0518-.0553&.0574-.0814\\
$3$&$1$&*&*&*&*&1            &.9971+.0626&.0489-.0601&.0531-.0862&.0598-.1119\\
   &$2$&*&*&*&*&*            &1            &.0492-.0915&.0546-.1172&.0623-.1424\\
\hline
   &$0$&*&*&*&*&*&*&1            &.9942+.0996&.9768+.1976    \\
$4$&$1$&*&*&*&*&*&*&*&1            &.9942+.0994    \\
   &$2$&*&*&*&*&*&*&*&*&1                \\
\hline
\multicolumn{2}{|c|}{$0$}
&\multicolumn{3}{|c|}{$l'=2$}
&\multicolumn{3}{|c|}{$l'=3$}
&\multicolumn{3}{|c|}{$l'=4$}\\
\hline
$l$&$N$&$N'=0$ &$N'=1$ &$N'=2$ &$N'=0$ &$N'=1$ &$N'=2$ &$N'=0$ &$N'=1$ &$N'=2$\\
\hline
   &$0$&1            &.9983+.0446&.9927+.0918&.0608-.0361&.0653-.0690&.0734-.1017&-.0039-.0010&-.0050-.0014&-.0069-.0020\\
$2$&$1$&*&1            &.9981+.0474&.0639-.0758&.0708-.1081&.0812-.1399&-.0061-.0031&-.0083-.0039&-.0110-.0052\\
   &$2$&*&*&1            &.0683-.1179&.0776-.1493&.0905-.1798&-.0103-.0063&-.0134-.0079&-.0169-.0100\\
\hline
   &$0$&*&*&*&1&.9934+.1025&.9733+.2043&.0487-.0335&.0547-.0616&.0660-.0886\\
$3$&$1$&*&*&*&*&1            &.9932+.1035&.0449-.0723&.0522-.1005&.0649-.1275\\
   &$2$&*&*&*&*&*            &1            &.0361-.1111&.0449-.1396&.0590-.1670\\
\hline
   &$0$&*&*&*&*&*&*&1&.9907+.1282&.9628+.2543\\
$4$&$1$&*&*&*&*&*&*&*&1            &.9906+.1289\\
   &$2$&*&*&*&*&*&*&*&*            &1            \\
\hline
\multicolumn{2}{|c|}{$-1$}
&\multicolumn{3}{|c|}{$l'=2$}
&\multicolumn{3}{|c|}{$l'=3$}
&\multicolumn{3}{|c|}{$l'=4$}\\
\hline
$l$&$N$&$N'=0$ &$N'=1$ &$N'=2$ &$N'=0$ &$N'=1$ &$N'=2$ &$N'=0$ &$N'=1$ &$N'=2$\\
\hline
   &$0$&1&.9948+.0935&.9769+.1954&.0547-.0378&.0618-.0697&.0760-.1007&-.0036-.0007&-.0045-.0008&-.0058-.0012\\
$2$&$1$&*&1            &.9936+.1032&.0544-.0832&.0641-.1146&.0808-.1450&-.0064-.0028&-.0083-.0035&-.0104-.0047\\
   &$2$&*&*            &1            &.0481-.1339&.0609-.1652&.0807-.1951&-.0126-.0056&-.0155-.0072&-.0185-.0096\\
\hline
   &$0$&*&*&*&1&.9891+.1376&.9553+.2752&.0440-.0338&.0517-.0606&.0667-.0857\\
$3$&$1$&*&*&*&*&1            &.9884+.1414&.0370-.0742&.0462-.1016&.0628-.1275\\
   &$2$&*&*&*&*&*            &1            &.0212-.1144&.0318-.1431&.0501-.1703\\
\hline
   &$0$&*&*&*&*&*&*&1&.9874+.1519&.9491+.3017\\
$4$&$1$&*&*&*&*&*&*&*&1            &.9870+.1542\\
   &$2$&*&*&*&*&*&*&*&*            &1            \\
\hline
\multicolumn{2}{|c|}{$-2$}
&\multicolumn{3}{|c|}{$l'=2$}
&\multicolumn{3}{|c|}{$l'=3$}
&\multicolumn{3}{|c|}{$l'=4$}\\
\hline
$l$&$N$&$N'=0$ &$N'=1$ &$N'=2$ &$N'=0$ &$N'=1$ &$N'=2$ &$N'=0$ &$N'=1$ &$N'=2$\\
\hline
   &$0$&1&.9891+.1430&.9478+.3089&.0398-.0308&.0475-.0554&.0641-.0797&-.0025-.0003&-.0030-.0001&-.0035-.0003\\
$2$&$1$&*&1            &.9844+.1704&.0374-.0709&.0473-.0955&.0662-.1196&-.0049-.0018&-.0060-.0021&-.0071-.0031\\
   &$2$&*&*            &1            &.0223-.1195&.0353-.1452&.0575-.1705&-.0110-.0036&-.0129-.0048&-.0147-.0070\\
\hline
   &$0$&*&*&*&1&.9847+.1686&.9355+.3406&.0363-.0303&.0445-.0535&.0609-.0748\\
$3$&$1$&*&*&*&*&1            &.9828+.1783&.0280-.0676&.0376-.0918&.0556-.1145\\
   &$2$&*&*&*&*&*            &1            &.0082-.1048&.0191-.1310&.0388-.1561\\
\hline
   &$0$&*&*&*&*&*&*&1&.9843+.1718&.9359+.3423\\
$4$&$1$&*&*&*&*&*&*&*&1            &.9834+.1766\\
   &$2$&*&*&*&*&*&*&*&*            &1            \\
\hline
\end{tabular}
\end{table}

\clearpage

\section{Higher-dimensional spheroidal harmonics}
\label{sec:ddim}

\subsection{Series solution}
\label{idaop}

Scalar $(n+4)$-dimensional spheroidal harmonics are defined as
solutions of the equation

\begin{eqnarray}
{1\over\sin\theta\cos^n\theta}\left({d\over d \theta} \sin\theta\cos^n\theta{dS_{kjm}\over d
\theta}\right) +\left[c^2\cos^2\theta -m^2\csc^2\theta -j(j+n-1)\sec^2\theta
+A_{kjm}\right]S_{kjm}=0\,. \label{ang}
\end{eqnarray}
In four dimensions there is only one possible rotation axis for a
cylindrically symmetric space-time, so there is only one angular
momentum parameter.  In higher dimensions there are several possible
choices of rotation axes. Therefore higher-dimensional black holes
will have, in general, several angular momentum parameters, each
referring to a particular rotation axis \cite{myersperry}.  Here we
focus on the simple case where there is {\it only one} rotation axis
(and correspondingly, only one angular momentum direction). The
angular separation constant $A_{kjm}$ now depends on three indices.
The integer $m=0,~\pm 1,~\pm 2\dots$ comes from separation of the
angle describing the azimuthal dependence of the perturbations around
the symmetry axis. Since (\ref{ang}) is invariant under the
replacement $m\to -m$ we will restrict our calculations to $m\geq
0$. The number of extra dimensions $n$ and the parameter
$j=0,~1,~2,\dots$ are related with the eigenvalues of hyperspherical
harmonics on the $n$-sphere, which are given by $-j(j+n-1)$.  The
parameter $k(=0,~1,~2\dots)$ plays the same role as $r/2$ in
(\ref{A,c=}) in the four-dimensional case, i.e. it labels the
eigenvalues for fixed values of $j$ and $m$.  Eq.~(\ref{ang})
describes the angular dependence of a scalar field perturbing a
$(n+4)$-dimensional Myers-Perry black hole with a single rotation
parameter, and is a simple generalization of the four-dimensional case
\cite{ida,frolov,vasudevan,css}.  When $n=1$ it also describes scalar
perturbations of smooth geometries in the D1-D5 stringy system
\cite{mathur}.
%
%

In the region of physical interest $\theta\in [0,\pi]$ the
differential equation (\ref{ang}) has three regular singular points:
$\theta=0,~\pi/2$ and $\pi$.  We make the following change in the angular
function:
\begin{eqnarray} \label{Frobenius}
S_{kjm}&=&\left(\sin\theta\right)^{\tilde{k}_0}\left(\cos\theta\right)^{\tilde{k}_1}y_{kjm}\,,
\end{eqnarray}
and using a Frobenius expansion for $y_{kjm}$ around $\theta=0,~\pi/2$
and $\pi$ we find the characteristic exponents: $\tilde{k}_0=|m|$,
$\tilde{k}_1=j$ or $\tilde{k}_1=1-n-j$.  The solution with
$\tilde{k}_1=1-n-j$ is irregular at $x=0$ in the higher-dimensional
case, so we take $\tilde{k}_1=j$.  Inserting the expansion
(\ref{Frobenius}) into the differential equation (\ref{ang}) with
$\tilde{k}_0=|m|$ and $\tilde{k}_1=j$, we find that the function
$y_{kjm}$ satisfies the differential equation
\begin{equation} \label{y eq}
\Bigg\{
(1-x^2){d^2\over d x^2}+
\left[\frac{(n+2j)(1-x^2)}{x}-2(|m|+1)x\right]{d\over d x}
+c^2x^2+\delta_{kjm}
\Bigg\}y_{kjm}=0\,,
\end{equation}
where $\delta_{kjm}\equiv A_{kjm}-j(j+n+1)-|m|(n+|m|+2j+1)$.

We assume a series expansion for $y_{kjm}$ such that
\begin{eqnarray}\label{HSH-Leaver}
S_{kjm}&=&(\sin\theta)^{|m|}(\cos\theta)^{j}\sum_{p=0}^\infty
\tilde{a}_p(\cos^2\theta)^p\,.
\end{eqnarray}
This series (if convergent) automatically satisfies the regularity
boundary conditions at $\theta=0,~\pi/2,~\pi$. Substituting
(\ref{HSH-Leaver}) into (\ref{ang}) we obtain the three-term recursion
relation \cite{css}
\begin{eqnarray}
\label{angrec}
&&\tilde \alpha_0\tilde{a}_1+\tilde \beta_0\tilde{a}_0=0\,,\\
&&\tilde \alpha_p\tilde{a}_{p+1}+\tilde \beta_p\tilde{a}_p
+\tilde \gamma_p\tilde{a}_{p-1}=0\,,\quad(p=1,2,\cdots)\nonumber
\end{eqnarray}
where
\begin{eqnarray}
\tilde \alpha_p&=&-2(p+1)(2j+n+2p+1)\,,\\
\tilde \beta_p&=&(j+|m|+2p)(j+n+|m|+2p+1)-A_{kjm}\,,\nonumber\\
\tilde \gamma_p&=&-c^2\,. \nonumber
\end{eqnarray}

The continued fraction equation for the separation constant has the
same form as Eq.~(\ref{cfeq}):
\begin{eqnarray}\label{cfeqD}
\tilde \beta_0-
{\tilde \alpha_0\tilde \gamma_1\over\tilde \beta_1-}
{\tilde \alpha_1\tilde \gamma_2\over\tilde \beta_2-}
{\tilde \alpha_2\tilde \gamma_3\over\tilde \beta_3-}
...=0 \,. \label{b-eq}
\end{eqnarray}

\subsection{Small-$c$ expansion}
\label{sec:ddimexp-small}

In this Section we provide an expansion of the eigenvalue $A_{kjm}$ in powers of $c$ around $c=0$. For $c=0$,
the eigenvalue $A_{kjm}$ is explicitly determined from the requirement that the series expansion has a finite
number of terms, since otherwise it is divergent \cite{leaver}. Imposing ${\tilde \beta_k}=0$ for some integer
$k\geq 0$ we have, for $c=0$ (an alternative derivation is presented in Appendix \ref{sec:A1}),
\be A_{kjm}=(2k+j+|m|)(2k+j+|m|+n+1)\,. \ee
If we identify
\be 2k=l-(j+|m|)\,, \ee
we can write $A_{kjm}=l(l+n+1)$. The integer $l$ is constrained to
satisfy the condition $l\geq (j+|m|)$, which is a simple
generalization of the four-dimensional case. An important difference
from the four-dimensional case is that now $l$ cannot be any positive
integer: for $k$ to be a positive or zero integer, only even (odd)
values of $l$ are admissible when $(j+|m|)$ is even (odd).

For finite but small $c$ we proceed as in Sec.~\ref{4Dsmallaw}. When
$c=0$ the recursion relation (\ref{b-eq}) has a finite number of terms
$k$. Therefore, expanding around $c=0$ we find it convenient (although
strictly not necessary) to use the $k$-th inversion of
Eq.~(\ref{b-eq}):
\be \tilde{\beta} _k-\frac{\tilde{\alpha}_{k-1}\tilde{\gamma} _{k}}{\tilde{\beta}
_{k-1}}\frac{\tilde{\alpha}_{k-2}\tilde{\gamma}_{k-1}}{\tilde{\beta}_{k-2}-}...\frac{\tilde{\alpha} _0 \tilde{\gamma} _1}{\tilde{\beta}_0}=\frac{\tilde{\alpha} _k
\tilde{\gamma} _{k+1}}{\tilde{\beta} _{k+1}-}\frac{\tilde{\alpha} _{k+1}\tilde{\gamma} _{k+2}}{\tilde{\beta}_{k+2}-}...\,\,\,\,\,\, \label{cont3}\ee
Now we expand the separation constant as a Taylor series:
\be
A_{kjm}=\sum_{p=0}^{\infty} \tilde{f}_pc^p\,,
\label{expansion2}
\ee
where
\begin{subequations}
\beq
\tilde{f}_0&=&l(l+n+1)\,, \\
\tilde{f}_1&=&0\,,\\
\tilde{f}_2&=&
\frac{-1+2j(j-1)+2l(l+1)-2m^2+2n(j+l)+n^2}{(2l+n-1)(2l+3+n)}\,,\\
\tilde{f}_3&=&0\,,\\
\tilde{f}_4&=& \frac{(j-l+|m|)(j+l-|m|+n-1)}{16(2l+n-3)(2l+n-1)^2}
\biggl [(2+j-l+|m|)(l+j-|m|+n-3)-\nonumber \\
& & 4(2l+n-3)\frac{-1+2j(j-1)+2l(l+1)-2m^2+2n(j+l)+n^2}{(2l+n-1)(2l+n+3)} \biggr ]-\nonumber \\
& & \frac{(j-l+|m|-2)(j+l+n-|m|+1)}{16(2l+5+n)(2l+n+3)^2}\biggl [ (j-l+|m|-4)(l+j+n-|m|+3)+\nonumber \\
& & 4(2l+5+n)\frac{-1+2j(j-1)+2l(l+1)-2m^2+2n(l+j)+n^2}{(2l+n-1)(2l+n+3)}\biggr ]\,,\\
\tilde{f}_5&=&0\,.
\eeq
\end{subequations}
The coefficients $\tilde f_1,\dots, \tilde f_5$ were obtained
substituting (\ref{expansion2}) into Eq.~(\ref{cont3}) and expanding
the resulting expression in powers of $c$.  Higher order coefficients
can be obtained easily, but the expressions are too lengthy to
reproduce them here. The four-dimensional results can be obtained by
setting $j=n=0$. They are in agreement with \cite{fackerell,seidel}
and with our Eqs.~(\ref{4dsmall1})-(\ref{4dsmall2}).

\subsection{Large and real $c$ (oblate case)}
\label{sec:ddimexp-large}

We now proceed to calculate the asymptotic behaviour of the oblate
HSHs (i.e., Eq. (\ref{ang}) with $c\in \mathbb{R}$) for large $c$,
following the same method as in Sec.~\ref{subsec:4SWSHlarge,real}.  In
order to find asymptotic solutions valid near the end-points
(i.e. ``inner solutions'') we make the change of independent variable
$u\equiv 2c(1-x)$.  The ordinary differential equation (\ref{y eq})
then becomes
\begin{equation} \label{eq:HSH,inner eq}
\begin{aligned}
&u\frac{d^{2}y_{kjm}}{du^{2}}+(|m|+1)\frac{dy_{kjm}}{du}
-\frac{1}{4}\left\{u -\frac{1}{c}\left(c^{2}+\delta_{kjm}\right)
\right\}y_{kjm}-
\\
&\quad-\frac{1}{4c}\left\{u^{2}\frac{d^{2}y_{kjm}}{du^{2}}+2(|m|+1)u
\frac{dy_{kjm}}{du}-
\frac{u^{2}}{4}y_{kjm}\right\}-\frac{(n+2j)u}{(2c-u)}
\left(1-\frac{u}{4c}\right)\frac{dy_{kjm}}{du}=0
\end{aligned}
\end{equation}
This equation is the same as the corresponding equation in four
dimensions where $\delta_{kjm}$ plays the role of $_{s}A_{lm}$, except
for the presence of the last term in Eq. (\ref{eq:HSH,inner eq}). This
last term can be neglected to leading order in $c$.  The asymptotic
solution of Eq. (\ref{eq:HSH,inner eq}) valid near the end-points is
therefore the same as the inner solution in four dimensions (given by
Eq. (3.10) of ~\cite{CO} with $h=0$), so that
$\delta_{kjm}=-c^2+2qc+{\cal O}(1)=A_{kjm}+{\cal O}(1)$
[cf. Eq.~(\ref{eq:series E for large w})].

Our task is now to determine the parameter $q$.  We can find an
asymptotic solution which is valid far from the end-points (the
``outer'' solution) as in \cite{CO}. The result is:
\begin{equation}\label{outd}
S_{kjm}^{\text{out}}(x)=(1-x^2)^{-1/2}x^{-n/2} \Big[a_{kjm}(1-x)^{q/2}(1+x)^{-q/2}e^{+cx}+
b_{kjm}(1-x)^{-q/2}(1+x)^{+q/2}e^{-cx}\Big]
\end{equation}
where $a_{kjm}$ and $b_{kjm}$ are constants of integration.  This
solution is similar to the outer solution in four dimensions, i.e.,
Eq. (3.26) of \cite{CO} with $h=0$. Unlike the four-dimensional case,
(\ref{outd}) is not valid very near the origin $x=0$.  Despite this
inconvenience, the matching of the inner and outer solutions in the
overlap region proceeds exactly as in four dimensions: the number of
extra dimensions $n$ plays no part in the matching.

The next step is to find the number of zeros of the HSHs. Equating
this number with the number of zeros of the asymptotic solution
determines the value of $q$.  Refs.~\cite{flammer,meixner} show that
scalar spheroidal harmonics (a special limit of HSHs with $s=n=j=0$)
are entire functions of $c$, so their number of zeros does not vary
with $c$. In particular, their number of zeros is the same as that of
the spherical harmonics ($s=n=j=c=0$), which is known.  Similarly,
Refs.~\cite{breuer,Hartle&Wilk'74,Stewart'75} show that SWSHs
($n=j=0$) are entire functions of $c$. Therefore their number of zeros
is easily obtained, since it does not vary with $c$.  The argument
used in both cases relies crucially on two properties:

\begin{itemize}

\item[(P1)] the fact that the indicial equation at any possible
regular singular point of the differential equation does not involve
$c$,

\item[(P2)] the continuity in $x$ of the coefficients in the
differential equation when the coefficient of the second order
derivative is equal to one.

\end{itemize}

We must be careful if we wish to use a similar argument to show that the number of zeros of the HSHs does not
vary with $c$.  First of all, the characteristic exponents ${\tilde \alpha_0}$ and ${\tilde \alpha_1}$ for the
HSH equation do not depend on $c$, and the differential equation (\ref{y eq}) does not possess any singular
points. Therefore the differential equations for $S_{kjm}$ and $y_{kjm}$ both satisfy the property (P1) above.
However, the coefficients of either differential equation are not continuous at $x=0$ when the coefficient of
the second order derivative is equal to one.  Therefore, property (P2) is not satisfied by either differential
equation. Nevertheless, we can still apply the argument separately to the regions $x\in (-1,0)$ and $x\in
(0,+1)$, where both properties (and other necessary ones) are satisfied by both differential equations.  Note
that no zero of $y_{kjm}$ can ``move'' from either region $x\in (-1,0)$ or $x\in (0,+1)$ to the other as $c$
varies, because this function cannot possess a zero at $x=0$.  We then conclude that the number of zeros of
$y_{kjm}$ in either interval $(-1,0)$ or $(0,+1)$ [and therefore in the whole region $(-1,+1)$] is independent
of $c$. In particular, it is equal to the number of zeros of $y_{kjm}$ in the corresponding region for $c=0$.
So the number of zeros of the HSHs for $x\in(-1,+1)$ for any real $c$ can be obtained from the special case
$c=0$.

\begin{figure}[hbt]
\centerline{
\includegraphics[width=9cm,angle=0]{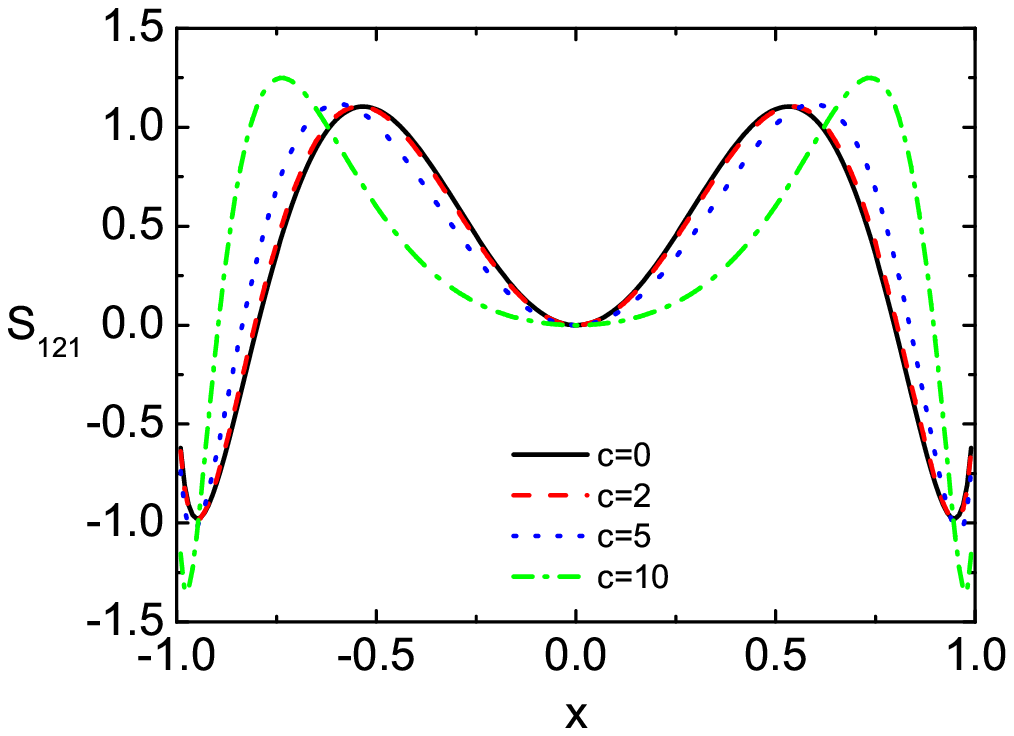}
\includegraphics[width=9cm,angle=0]{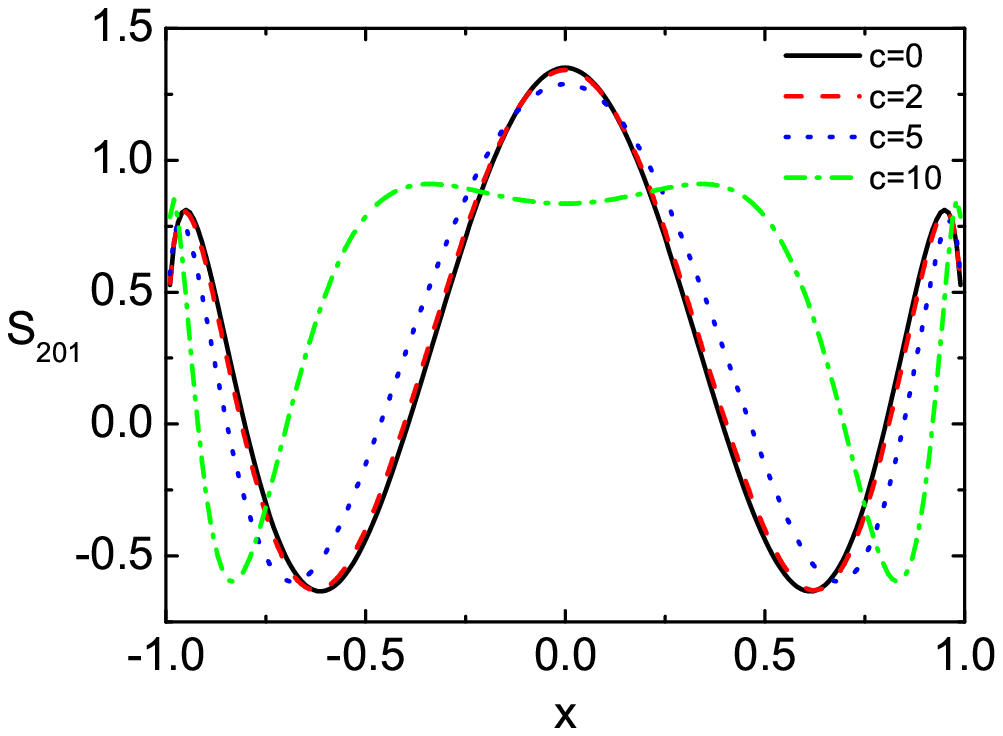}}
\caption{Some representative oblate eigenfunctions $S_{kjm}$ for
different values of $c$. The plot on the left refers to $n=2$. For
$c=0$ the eigenfunction can be obtained analytically (see Appendix
\ref{sec:A1}) with the result
$S_{121}=\left[45(1-x^2)/32\right]^{1/2}\left (7-11x^2\right)x^2$.
In the plot on the right $n=1$, and for $c=0$ we have
$S_{201}=\left[693(1-x^2)/380\right]^{1/2} \left(1-8 x^2+10
x^4\right)$.  }
\label{fig:HighDS}
\end{figure}

In Appendix \ref{sec:A1} we find an exact solution of the HSH equation
for $c=0$ and show that the number of zeros of the HSHs in $x\in
(-1,+1)$, excluding $x=0$, is equal to $2k$. The HSHs possess a zero
at $x=0$ if $j>0$ and do not if $j=0$; however, we will not need this
property. As a check of the analytic results we performed numerical
calculations of the eigenfunctions using Eq.~(\ref{HSH-Leaver}); some
of these numerically computed, normalized eigenfunctions are shown in
Fig.~\ref{fig:HighDS}.

The matching of the number of zeros of the HSH with that of the
asymptotic solution proceeds now as in four dimensions. Because of the
forms of the inner and outer asymptotic solutions, the number of zeros
of the asymptotic solution (excluding $x=0$) has the dependence on $q$
given in Eq. (4.4) of \cite{CO} with $h=0$ ($h$ in that paper is what
we call $s$ here).  We can equate this number to the number of zeros
of the HSH (excluding possible zeros at $x=0$), which is $2k=l-|m|-j$.

The resulting equation is the same as Eq. (4.4) in \cite{CO} with
$h=0$, and with $z_0$ replaced by $j$.  The value of the parameter $q$
is therefore
\begin{equation}\label{qfirst}
q=l-j+1=2k+|m|+1\,.
\end{equation}
The asymptotic behaviour of the separation variable $A_{kjm}$ is then given by
\begin{equation} \label{eq:higher dim. A,large c}
A_{kjm}=-c^2+2(l-j+1)c+{\cal O}(1)\,.
\end{equation}
There is an imprint of the extra dimensions (through the parameter
$j$) already at order $O(c)$.
Notice that the above asymptotic expansion (\ref{eq:higher
dim. A,large c}) does not agree in the limit $n=j=0$ with the
four-dimensional result, Eqs.~(\ref{eq:series E for large w}) and
(\ref{eq:val. of q}) with $s=0$.  The reason is that the present
analysis is based on picking the characteristic exponent $\tilde{k}_1=j$
while disregarding $\tilde{k}_1=1-n-j$ in Eq. (\ref{Frobenius}).  The
latter value should {\it not} be disregarded when $n=j=0$, in which
case $\alpha_1$ can take on both values, 0 and 1.

\begin{figure}[hbt]
\centerline{
\includegraphics[width=9cm,angle=0]{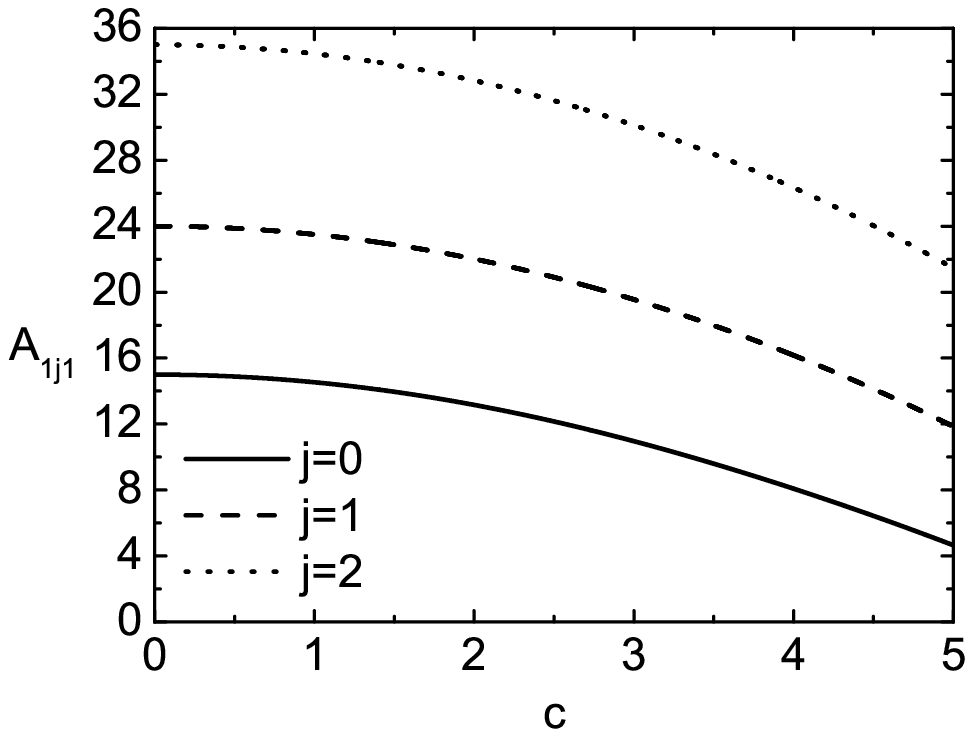}
\includegraphics[width=9cm,angle=0]{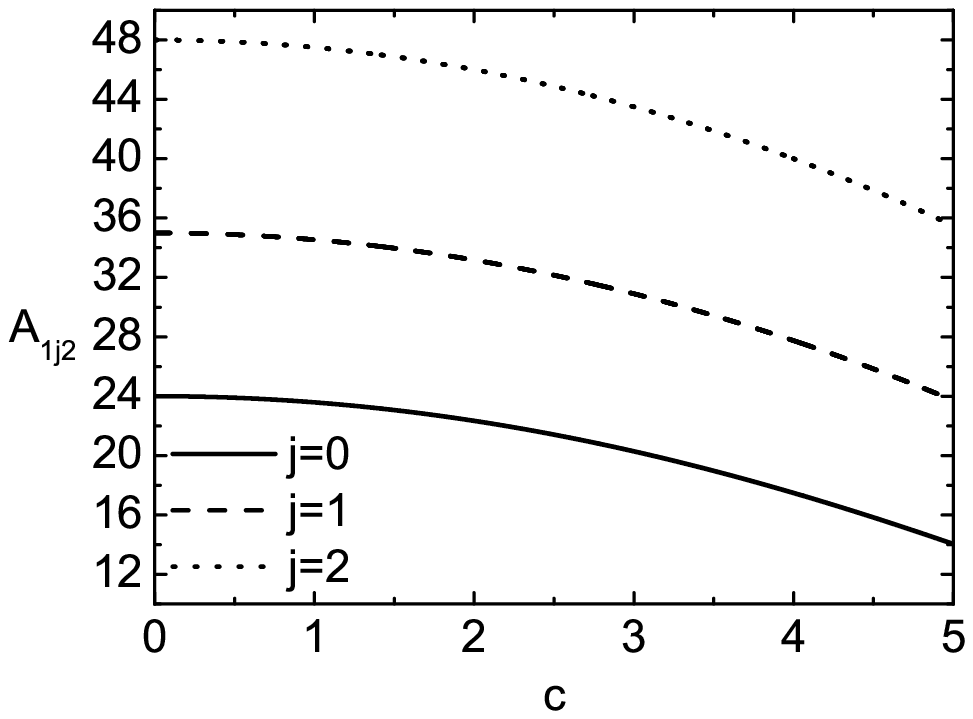}}
\caption{Oblate eigenvalues of the HSH with $k=n=1$ for $m=1$ (left),
$m=2$ (right), and different values of $j$.}
\label{fig:HSWSH}
\end{figure}

To conclude this Section, we back up our analytic results by numerical
calculations of the oblate eigenvalues obtained solving
Eq.~(\ref{b-eq}). In Fig.~\ref{fig:HSWSH} we plot the eigenvalues for
$n=k=1$, $m=1,~2$ and $j=0,~1,~2$; these values of $(n,k,j,m)$ have
been chosen to match those in Figs. 2-7 of \cite{ida}. We checked that
the asymptotic behavior of the oblate eigenvalues, both in the limit
$c\to 0$ and in the limit $c\to \infty$, is in agreement with the
analytic predictions.

\subsection{Large and pure-imaginary $c$ (prolate case)}
\label{sec:ddimexp-large,prolate}

In this Section we complete the picture computing a series expansions
of the eigenvalue $A_{kjm}$ for large and pure-imaginary values of
$c$. The present analysis is similar to the one in
Sec.~\ref{prolate-large}.

We first find an approximation valid near the origin when $|c_I|\to \infty$ by
introducing a new independent variable $u\equiv\sqrt{2|c_I|}x$ and using
again the transformation (\ref{Frobenius}) (with $\tilde{k}_0=|m|$ and
$\tilde{k}_1=j$).  The resulting ordinary differential equation for
the function $y_{kjm}$ is the same as Eq.~(\ref{y eq}), but in terms
of the new variable $u$:
\begin{equation} \label{ODEy(u),HSH,prolate}
\Bigg\{
\left(2|c_I|-u^2\right){d^2\over d u^2}+
\left[\frac{2|c_I|(n+2j)}{u}-\left(n+2j+2|m|+2\right)u\right]{d\over d u}
+\delta_{kjm}-\frac{|c_I|u^2}{2}
\Bigg\}y_{kjm}=0\,.
\end{equation}
%
Now we apply the same approximation as in Sec.~\ref{prolate-large}.
In the limit $|c_I|\to \infty$ and in the inner region $1/|c_I|\ll |x| \ll
1$, Eq.~(\ref{ODEy(u),HSH,prolate}) reduces to
\begin{equation} \label{ODEy(u),HSH,prolate,inner}
\Bigg\{
{d^2\over d u^2}+\frac{(n+2j)}{u}{d\over d u}
+q-\frac{u^2}{4}
\Bigg\}y^{\text{inn}}_{kjm}=0\,,
\end{equation}
where we have set $\delta_{kjm}\sim 2q|c_I|+{\cal
O}\left(|c_I|^0\right)$, and $q$ is an unknown parameter that we shall
now determine, not to be confused with the parameter introduced in
(\ref{qfirst}).  With respect to the four-dimensional case,
Eq.~(\ref{ODEy(u),HSH,prolate,inner}) contains an extra term with a
first derivative.  The presence of this term means that the inner
solution $y^{\text{inn}}_{kjm}$ is not a parabolic cylinder function,
as in the four-dimensional case.  By applying the change of variables
\begin{equation}
y^{\text{inn}}_{kjm}(u)=u^{-(n+2j)/2}v(u)\,,
\end{equation}
Eq.~(\ref{ODEy(u),HSH,prolate,inner}) becomes
\begin{equation} \label{ODEy(u),HSH,prolate,inner2}
\Bigg\{
{d^2\over d u^2}+q-\frac{(n+2j)(n+2j-2)}{4u^2}
-\frac{u^2}{4}
\Bigg\}v=0\,.
\end{equation}
The solution of this equation can be expressed in terms of a Laguerre
polynomial $L^{(\tilde{\alpha})}_{\tilde{n}}$ \cite{szego}.  It can be
checked that the asymptotic approximation of the prolate HSHs valid in
the inner region $1/|c_I|\ll |x| \ll 1$ as $|c_I|\to \infty$, which is
regular at $x=0$, is
\begin{equation}
y^{\text{inn}}_{kjm}(u)=e^{-u^2/4}L^{(\tilde{\alpha})}_{\tilde{n}}\left(u^2/2\right)\,,
\end{equation}
where
\begin{equation}
\tilde{\alpha}=\frac{n+2j-1}{2}
\qquad \text{and} \qquad
\tilde{n}=\frac{2q-(n+2j+1)}{4}\,.
\end{equation}
The number of positive zeros of the Laguerre polynomial
$L^{(\tilde{\alpha})}_{\tilde{n}}(x)$ is equal to $\tilde{n}$ if (as
in our case) $\tilde{\alpha}>-1$ \cite{szego}.  It can easily be
checked that all these zeros lie within the inner region.  The number
of zeros of $y^{\text{inn}}_{kjm}$ in the inner region [or
equivalently in the region $x\in (-1,+1)$] is therefore equal to
$2\tilde{n}$.

For $|c_I|\to\infty$, an outer solution which is a valid approximation
of the prolate HSH near the end-points can be obtained using a
WKB-type expansion, as in \cite{CO} and in
Sec.~\ref{prolate-large}. The result is
\begin{equation}
\begin{aligned}\label{outd-end}
S_{kjm}^{\text{out}}(x)=
&
a_{kjm} \left(1-x^2\right)^{-1/4} x^{-1/2+q-n}
\left(1+\sqrt{1-x^2}\right)^{-q}e^{+|c_I|\sqrt{1-x^2}}+
\\ &
b_{kjm} \left(1-x^2\right)^{-1/4} x^{-1/2-q-n}
\left(1+\sqrt{1-x^2}\right)^{q}e^{-|c_I|\sqrt{1-x^2}}
\end{aligned}
\end{equation}
where $a_{kjm}$ and $b_{kjm}$ are constants of integration.  For
$n=j=0$, (\ref{outd-end}) reduces to the corresponding outer solution
(\ref{outer,prolateSWSH}) in the prolate SWSH case if $s=0$ and we
replace $L+1/2$ there by $q$.  Furthermore (\ref{outd-end}) is valid
in the same region as (\ref{outer,prolateSWSH}): for
$|x|>>1/\sqrt{|c_I|}$ and $|x|\sim 1-O(|c_I|^\epsilon)$, with
$-1<\epsilon\leq 0$.  Matching the outer solution (\ref{outd-end})
with the inner solution in the overlap region we find
\begin{equation}
S_{kjm}^{\text{out}}(x)=
\frac{(-1)^{\tilde{n}}2^{q}(\tilde{\alpha}+1)!\tilde{\alpha}!}
{(\tilde{\alpha}+\tilde{n}-2)!\tilde{n}!(\tilde{\alpha}+\tilde{n})!}
|c_I|^{\tilde{n}}e^{-3|c_I|/2} \left(1-x^2\right)^{-1/4} x^{-1/2+q-n}
\left(1+\sqrt{1-x^2}\right)^{-q}e^{+|c_I|\sqrt{1-x^2}}\,.
\end{equation}
This outer solution has no zeros. Like in the prolate SWSH case with
$s=0$, all zeros of the asymptotic solution for $|c_I|\to\infty$ are
provided by the inner solution.  We want to equate their number to the
number of zeros of the exact solution $y_{kjm}$ in the region $x\in
(-1,+1)$, which is given in Appendix \ref{sec:A1} for $c=0$, is
independent of $c^2\in (-\infty,+\infty)$ and is equal to
$2k$. Therefore $2q=4k+n+2j+1$, and the separation constant for large
$|c_I|$ is
\begin{equation}
\label{prold}
A_{kjm}=\delta_{kjm}+{\cal O}\left(|c_I|^0\right)=
(4k+n+2j+1)|c_I|+{\cal O}\left(|c_I|^0\right)=
\left[2(l-|m|)+n+1\right] |c_I|+{\cal O}\left(|c_I|^0\right)\,.
\end{equation}
\begin{figure}[hbt]
\centerline{
\includegraphics[width=9cm,angle=0]{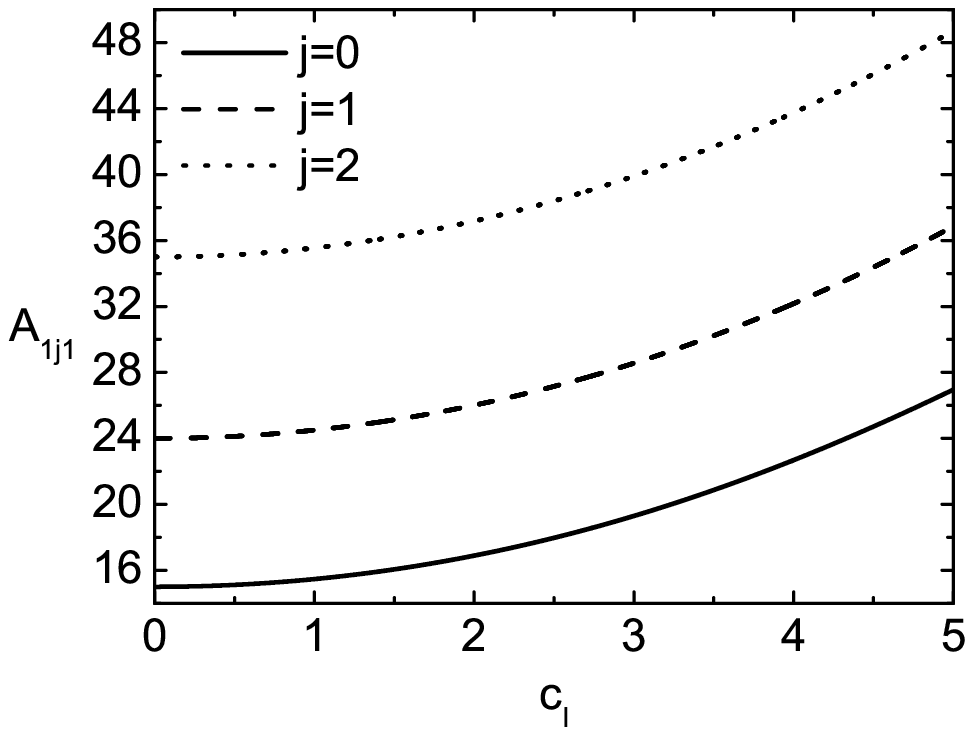}
\includegraphics[width=9cm,angle=0]{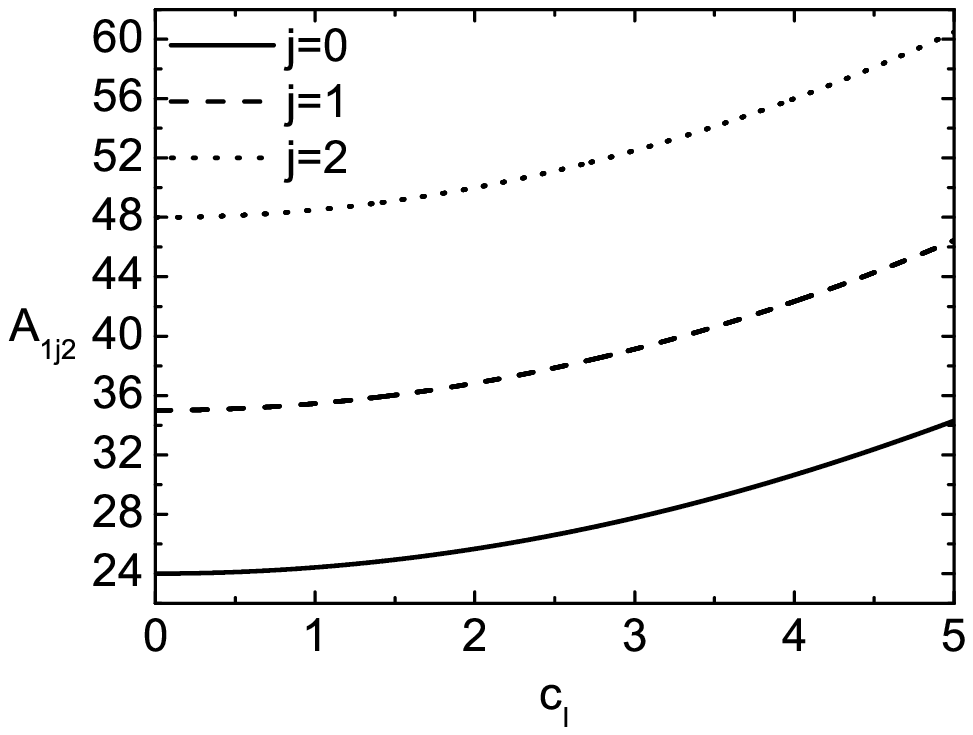}}
\caption{Prolate eigenvalues of the HSH with $k=n=1$ for $m=1$ (left),
$m=2$ (right), and different values of $j$.}
\label{fig:HSWSHp}
\end{figure}

The expansion (\ref{prold}) is in agreement with the eigenvalues
obtained from a numerical solution of Eq.~(\ref{cfeqD}) in the prolate
case, some of which are plotted in Fig.~\ref{fig:HSWSHp}.

\section{Conclusions}

In this paper we provided a complete numerical and analytic analysis
of four-dimensional spin-weighted spheroidal harmonics (SWSHs) and of
higher-dimensional spheroidal harmonics (HSHs). For reference we list
below what we regard as the most useful results, be they known in the
existing literature or originally derived in this paper.

For SWSHs, the main results are:

\begin{itemize}
\item[1)] Eq.~(\ref{leavers}) gives a series representation of the
SWSHs, and a simple computational algorithm to determine eigenvalues
and eigenfunctions by recursion;

\item[2)] Eq.~(\ref{expansion}) is the series expansion of the
eigenvalues for small $c$;

\item[3)] Eq.~(\ref{eq:series E for large w}) is the series expansion
of the eigenvalues for large $c\in \mathbb{R}$ (oblate case);

\item[4)] Eq.~(\ref{s0asy}) (for spin 0) and Eq.~(\ref{alm}) (for
general $s$, but to lower order) are series expansions of the
eigenvalues for large $c$ when $\ii c\in \mathbb{R}$ (prolate case);

\item[5)] Eq.~(\ref{Slm-orth}) and, more accurately,
Eq.~(\ref{swshsp}) provide analytic approximations of the scalar
products between SWSHs corresponding to slowly damped eigenfrequencies
of Kerr black holes.
\end{itemize}

For HSHs, the main results are:

\begin{itemize}
\item[1)] Eq.~(\ref{HSH-Leaver}) gives a series representation of the
HSHs, and a simple computational algorithm to determine eigenvalues
and eigenfunctions by recursion;

\item[2)] Eq.~(\ref{expansion2}) is the series expansion of the
eigenvalues for small $c$;

\item[3)] Eq.~(\ref{eq:higher dim. A,large c}) is the series expansion
of the eigenvalues for large $c\in \mathbb{R}$ (oblate case);

\item[4)] Eq.~(\ref{prold}) is the series expansion of the eigenvalues
for large $c$ when $\ii c\in \mathbb{R}$ (prolate case);

\item[5)] Eq.~(\ref{finaljacobi}) is an {\it exact} solution for the
eigenfunctions of the HSH equation for $c=0$ in terms of Jacobi
polynomials; Appendix~\ref{sec:A2} contains another exact solution of
the HSH equation for special values of the arguments.
\end{itemize}

In our opinion, the main unsolved theoretical problem is to clarify
why Eq.~(\ref{alm}) is in good agreement with the numerics for
$|s|\neq 0$, even though the number of zeros of the eigenfunction
seems to depend on $c$ when $c$ is not real. Given the excellent
accuracy of Eq.~(\ref{swshsp}) in the four-dimensional case, it would
be desirable to extend the perturbative formalism developed by Press
and Teukolsky \cite{PT} to higher dimension. It could also be useful
to push the expansions (\ref{alm}), (\ref{eq:higher dim. A,large c})
and (\ref{prold}) to higher order, and to extend the present
understanding of the location of the spheroidal branch points
\cite{barrowes} to the case $s\neq 0$.  Finally, it would be an
exciting and interesting challenge to extend our results for
higher-dimensional {\it scalar} harmonics to higher-dimensional
harmonics with $s\neq 0$.  We leave these issues for future work.

\section*{Acknowledgements}
V.~C. acknowledges financial support from FCT through PRAXIS XXI
programme, and from Funda\c c\~ao Calouste Gulbenkian through Programa
Gulbenkian de Est\'{\i}mulo \`a Investiga\c c\~ao
Cient\'{\i}fica. This work was supported in part by the National
Science Foundation under grant PHY 03-53180.  M.~C. is delighted to
thank Ted Cox for his help with numerous and humorous comments, and
the Cosmogrid consortium for the use of computational resources.  
We are very grateful to Sam Dolan for correcting some errors in
Eq.~(2.16).

\appendix
\section{Eigenvalues, eigenfunctions and zeros of the HSH equation for $c=0$}
\label{sec:A1}
In the $c=0$ limit the HSH equation can be solved exactly, and the
eigenvalues can be computed analytically. The general solution for the
eigenvalues was first found by Ida {\it et al.} \cite{ida}. However,
their method relies on the continued fraction representation and does
not provide an explicit analytic expression for the
eigenfunctions. The purpose of this Appendix is to compute explicitly
the eigenfunctions and eigenvalues for $c=0$.

Let us introduce a new wavefunction $\Psi$:
\be S_{kjm}=(\sin{\theta})^{|m|}(\cos\theta)^{|j|}\Psi\,.\ee
Setting $c=0$ and $z\equiv (\cos\theta) ^2$ in (\ref{ang}) we get
\be z(1-z)\partial^2_z\Psi+\left (j+\frac{n+1}{2}-\left(\frac{n+3}{2}+j+|m|\right)z \right
)\partial_z\Psi-\left ((j+m)(j+|m|+n+1)/4-A_{kjm}/4\right )\Psi=0\,.\ee
This is an hypergeometric equation:
\be z(1-z)\partial^2_z\Psi+\left (\gamma-(\alpha+\beta+1)z \right )\partial_z\Psi-\left (\alpha \beta\right
)\Psi=0\,, \label{h1}\ee
with
\be \gamma=j+\frac{n+1}{2}\,, \quad \alpha=\frac{1}{2}\left (j+|m|+\frac{n+1}{2}+\sqrt{A_{kjm}
+\left(\frac{n+1}{2}\right )^2}\right )\,, \quad \beta=\frac{1}{2} \left
(j+|m|+\frac{n+1}{2}-\sqrt{A_{kjm}+\left (\frac{n+1}{2}\right )^2}\right )\,. \ee
The general solution for $z\in(0,1)$ can be written as
\be \Psi=A z^{1-\gamma}F[\alpha-\gamma+1,\,\beta-\gamma+1,\,2-\gamma,z]+BF[\alpha,\,\beta,\,\gamma,z]\,. \ee
For small $z$ ($\sin\theta \simeq 1$) $F\simeq 1$, and we get
\be
\Psi \simeq A\cos\theta^{-2|j|}+B \,,\quad
S_{kjm} \simeq A\cos\theta^{-|j|}+B\,.
\ee
Regularity requires $A=0$. To analyse the point $z=1$ we use the property
\be
F[a,b,c,z]=(1-z)^{c-a-b}
\frac{\Gamma[c]\Gamma[a+b-c]}{\Gamma[a]\Gamma[b]}F[c-a,c-b,c-a-b+1,1-z]+
\frac{\Gamma[c]\Gamma[c-a-b]}{\Gamma[c-a]\Gamma[c-b]}F[a,b,a+b-c+1,1-z]\,.
\label{fe2}
\ee
So for $z\simeq 1$
\be
\Psi \simeq
(\sin\theta)^{-2|m|}
\frac{\Gamma[\gamma]\Gamma[\alpha+\beta-\gamma]}{\Gamma[\alpha]\Gamma[\beta]}+
\frac{\Gamma[\gamma]\Gamma[\gamma-\alpha-\beta]}
{\Gamma[\gamma-\alpha]\Gamma[\gamma-\beta]}\,,
\ee
and requiring regularity of the wavefunction we get $\alpha=-k$ or
$\beta=-k$ ($k=0\,,1\,,2\,,\dots$). This leads to
\be A_{kjm}=\left (2k+j+|m|\right ) \left (2k+j+|m|+n+1\right )\,,\ee
in agreement with the result found by Frolov and Stojkovic for $n=1$
\cite{frolov} and by Ida {\it et al.} for general $n$ \cite{ida}. The
eigenfunctions can be written in terms of hypergeometric functions:
\be\label{EFs}
S_{kjm}=(\sin{\theta})^{|m|}(\cos\theta)^{|j|}
F[\alpha,\,\beta,\,\gamma,cos^2\theta] \,,
\ee
For reference we list some eigenfunctions for $n=1$, obtained from
(\ref{EFs}) and then normalized:
\beq
S_{000}&=&(1/2)^{1/2}\,,~
S_{010}=(3/2)^{1/2}\cos\theta\,,~
S_{001}=(3/4)^{1/2}\sin\theta\,,\nonumber\\
S_{011}&=&(15/4)^{1/2}\cos\theta\sin\theta\,,~
S_{021}=(35/4)^{1/2}\cos^2\theta\sin\theta\,,\nonumber\\
S_{100}&=&(15/14)^{1/2}\left(1-2\cos^2\theta\right)\,,~
S_{101}=(21/16)^{1/2}
\left(1-3\cos^2\theta\right)\sin\theta\,,\nonumber\\
S_{111}&=&(63/4)^{1/2}\cos\theta
\left(1-2\cos^2\theta\right)\sin\theta\,,~
S_{121}=(31185/464)^{1/2}\cos^2\theta
\left(3-5\cos^2\theta\right)\sin\theta\,,\nonumber\\
S_{102}&=&(315/208)^{1/2}\sin^2\theta
\left(1-4\cos^2\theta\right)\,,~
S_{112}=(3465/592)^{1/2}\cos\theta
\left(2-5\cos^2\theta\right)\sin^2\theta\,,\nonumber\\
S_{122}&=&(45045/368)^{1/2}\cos^2\theta
\left(1-2\cos^2\theta\right)\sin^2\theta\,,~
S_{202}=(45045/22208)^{1/2}\left[\sin^2\theta
\left(1-10\cos^2\theta+15\cos^4\theta\right)\right]
\,.\nonumber
\eeq
%
To convert hypergeometric functions to Jacobi polynomials ${\cal
P}_n^{\alpha\,,\beta}(z)$ we can use the property \cite{AS}:
\be F[a,-n,c,z]=\frac{n!}{(c)_n}{\cal P}_n^{c-1\,,a-c-n}(1-2z)\,,\ee
where $(c)_n \equiv \Gamma[c+n]/\Gamma[c]$ is the Pochhammer symbol.
As a result, the (non normalized) wavefunctions can be expressed as
\be
S=(\sin{\theta})^{|m|}(\cos\theta)^{|j|}
{\cal P}_k^{\alpha\,,\beta}(1-2\cos^2\theta)\,,
\label{finaljacobi}
\ee
where $\alpha=(k-1)/2+j$ and $\beta=|m|$.

We can now use this analytic expression of the eigenfunctions to
determine the number of zeros in the interval of physical
interest. Define Klein's symbol
\begin{equation}
E(u)=\left\{ \begin{array}{ll}
             0   & \mbox{if $u\leqslant0$}\\
             $[$u$]$    & \mbox{if $u$}\, {\rm positive \,and\,nonintegral}\\
             u-1  &\mbox{if $u=1\,,2\,,...$}\,,
\end{array}\right.
\label{Klein}
\end{equation}
where $[\,]$ denotes the floor function, and
\be X(\alpha\,,\beta)=E\left (\frac{1}{2}(|2n+\alpha+\beta+1|-|\alpha|-|\beta|+1) \right )\,. \ee
Then the number of zeros $N(\alpha\,,\beta)$ of ${\cal
P}_n^{\alpha\,,\beta}(z)$ in the interval $[-1,1]$ is given by
\cite{szego}
\begin{equation}
N(\alpha\,,\beta)=\left\{ \begin{array}{ll}
             2$[(X+1)/2]$   & \mbox{for $(-1)^n\left(%
\begin{array}{c}
  n +\alpha\\
  \alpha \\
\end{array}%
\right)\left(%
\begin{array}{c}
  n +\beta\\
  \beta \\
\end{array}%
\right)>0$}\,,
\\
             $2[X/2]+1$    & \mbox{for $(-1)^n\left(%
\begin{array}{c}
  n +\alpha\\
  \alpha \\
\end{array}%
\right)\left(%
\begin{array}{c}
  n +\beta\\
  \beta \\
\end{array}%
\right)>0$}\,.
\end{array}\right.
\end{equation}
Specializing to our case, we get that the number of zeros of ${\cal
P}_k^{\alpha\,,\beta}(z)$ in the $(-1\,,1)$ interval is $k$. But
${\cal P}_k^{\alpha\,,\beta}(1-2\cos^2\theta)={\cal
P}_k^{\alpha\,,\beta}(-\cos2\theta)$, so in the interval $(0\,,\pi)$
we have $2k$ zeros.  We know from Eq. (\ref{Frobenius}) that $y_{lm}$
does not possess a zero at $\cos\theta=0,\pm 1$.  This is ratified
here by the fact that ${\cal P}_k^{\alpha\,,\beta}(-\cos2\theta)$ does
not possess a zero at $\cos\theta=0$ nor at $\cos\theta=\pm 1$ because
neither $\alpha$ nor $\beta$ can take on the values $-1,-2,\ldots ,-k$
when $k,~j,~|m|\ge 0$.

\section{Solution of the HSH equation for special values of the arguments}
\label{sec:A2}

It is well known that the four-dimensional scalar spheroidal harmonics
admit an exact solution when $m=1$ and $_{0}A_{l1}=-c^2$
\cite{flammer}. For these values the spheroidal equation reduces to
\be
\partial^2_x\Psi -c^2 \Psi=0\,,
\ee
where $\Psi=(x^2-1)^{1/2}\,_{0}S_{l1}$. Then the solution is
elementary: $\Psi=\sin (\ii c x)$, with
\be
\ii c=l\pi\,, \quad l=0,1,2... \label{partw}
\ee
In an attempt to generalize this simple particular solution to the
$(n+4)$-dimensional case, we introduce a wavefunction
\be \bar \Psi=x^{n/2}\sqrt{x^2-1}\,S_{kjm}\,. \label{transfpart}\ee
We find that $\bar \Psi$ admits the same elementary solution as in the
four-dimensional case only if $j=0$, $m=1$, $n=2$, $A_{k01}=-c^2-2$,
and then $c$ takes the value (\ref{partw}). We could not find other
values of the arguments leading to such a simple solution.


\end{document}